\documentclass[sigconf,nonacm]{template/acmart}
\usepackage[utf8]{inputenc}
\usepackage{todonotes}
\usepackage{booktabs}
\usepackage{amsthm}
\usepackage{tabularx}

\usepackage[markup=underlined,final]{changes} %
\usepackage{tikz}
\usetikzlibrary{positioning,shadows,fit,calc,shapes,backgrounds,arrows.meta}
\usepackage{multirow}
\usepackage{makecell}
\usepackage{listings}
\usepackage[inline]{enumitem}
\usepackage{microtype} %
\PassOptionsToPackage{breakurls}{hyperref}
\usepackage{url}
\usepackage{soul}
\usepackage{stfloats}
\usepackage{comment}
\usepackage{rotating}
\usepackage{xspace}
\usepackage{mathtools}
\usepackage{comment}
\usepackage{subcaption}

\usepackage{colortbl}

\usepackage{fontawesome5}

\usepackage{csquotes} %
\usepackage{cleveref}
\renewcommand{\autoref}{\Cref}
\Crefname{section}{Sec.}{Sec.}
\crefname{figure}{Fig.}{Fig.}
\Crefname{equation}{Eq.}{Eq.}

\usepackage{placeins}
\usepackage{lipsum}
\usepackage{xparse} %
\usepackage{mleftright}\mleftright %
\usepackage{mathpartir}

\usepackage{color}
\definecolor{keywordcolor}{rgb}{0.7, 0.1, 0.1}   %
\definecolor{tacticcolor}{rgb}{0.0, 0.1, 0.6}    %
\definecolor{commentcolor}{rgb}{0.25, 0.5, 0.35}  %
\definecolor{symbolcolor}{rgb}{0.0, 0.1, 0.6}    %
\definecolor{sortcolor}{rgb}{0.1, 0.5, 0.1}      %
\definecolor{attributecolor}{rgb}{0.7, 0.1, 0.1} %
\definecolor{stringcolor}{rgb}{0.5, 0.15, 0.55}   %

\usepackage{bbm} %

\newtheorem{innerdef}{Definition}

\crefname{innerdef}{def.}{defs.}

\newtheorem*{innerdef*}{Definition}
\newenvironment{mdef*}{\begin{innerdef*}}{\end{innerdef*}}

\newtheorem{innernotation}{Notations}

\newtheorem*{innernotations*}{Notations}
\newenvironment{mnotation*}{\begin{innernotations*}}{\end{innernotations*}}

\newtheorem{innerremark}{Remark}
\newenvironment{mrmk}{\begin{innerremark}}{\end{innerremark}}

\crefname{innerremark}{rmk.}{rmks.}

\newtheorem*{innerremark*}{Remark}
\newenvironment{mrmk*}{\begin{innerremark*}}{\end{innerremark*}}

\newtheorem{innerexample}{Example}
\newenvironment{mex}{\begin{innerexample}}{\end{innerexample}}

\crefname{innerexample}{ex.}{ex.}

\newtheorem*{innerexample*}{Example}
\newenvironment{mex*}{\begin{innerexample*}}{\end{innerexample*}}

\newtheorem{innerlem}{Lemma}
\newenvironment{mlemma}{\begin{innerlem}}{\end{innerlem}}

\crefname{innerlem}{lem.}{lems.}

\newtheorem{innercor}{Corollary}

\crefname{innercor}{corollary}{corollaries}

\newtheorem{innerth}{Theorem}

\crefname{innerth}{th.}{ths.}

\newtheorem*{innerth*}{Theorem}
\newenvironment{mth*}[1][]{\begin{innerth*}[#1]}{\end{innerth*}}

\newtheorem{innerprop}{Property}
\newenvironment{mprop}{\begin{innerprop}}{\end{innerprop}}

\crefname{innerprop}{prop.}{props.}

\newtheorem*{innerprop*}{Property}
\newenvironment{mprop*}{\begin{innerprop*}}{\end{innerprop*}}

\def\bitcoinA{%
  \leavevmode
  \vtop{\offinterlineskip %
    \setbox0=\hbox{B}%
    \setbox2=\hbox to\wd0{\hfil\hskip-.03em
    \vrule height .3ex width .15ex\hskip .08em
    \vrule height .3ex width .15ex\hfil}
    \vbox{\copy2\box0}\box2}}

\theoremstyle{plain}

\newcommand{\etal}{et al.}

\definechangesauthor[color=violet]{LV}
\definechangesauthor[color=blue]{MS}
\definechangesauthor[color=green]{SJ}
\definechangesauthor[color=red]{MM}
\setcommentmarkup{\todo[color={authorcolor!20},size=\tiny]{#3: #1}}

\newcommand{\code}[1]{\texttt{\hyphenchar\font=`\-\small#1}}

\newcommand{\CR}[1]{}

\newcommand{\bfparagraph}[1]{\smallskip\noindent\textbf{#1}.\,}

\newcommand{\attack}[2]{\medskip\noindent\textbf{\csname fa#1\endcsname\, #2}\\}

\newcommand{\cir}[1]{\tikz[baseline=(char.base)]{%
    \node[anchor=base, draw, fill=white, font=\bf\scriptsize, circle, inner sep=0.2pt, minimum width=0.8em] (char) {#1};}}

\newcommand{\cirb}[1]{\tikz[baseline=(char.base)]{%
    \node[anchor=base, draw, fill=black, text=white, font=\bf\scriptsize, circle, inner sep=0.2pt, minimum width=0.8em] (char) {#1};}}

\newcommand{\meventStyle}[1]{\tikz[baseline=(char.base)]{%
    \node[anchor=base,draw,color=gray,text=black, fill=gray!30, font=\scriptsize,rounded corners, inner sep=1.5pt, minimum height=1em, minimum width=0.8em] (char) {#1};}}

\newcounter{msg}

\newcounter{group}

\crefformat{msg}{msg.~#2\cir{#1}#3}
\Crefformat{msg}{Message~#2\cir{#1}#3}
\crefmultiformat{msg}{msg.~#2\cir{#1}#3}{ and~#2\cir{#1}#3}{, #2\cir{#1}#3}{ and~#2\cir{#1}#3}
\Crefmultiformat{msg}{Messages~#2\cir{#1}#3}{ and~#2\cir{#1}#3}{, #2\cir{#1}#3}{ and~#2\cir{#1}#3}
\crefrangeformat{msg}{msg.~#2\cir{#1}#3 to~#2\cir{#1}#3}
\Crefrangeformat{msg}{Messages~#2\cir{#1}#3 to~#2\cir{#1}#3}

\crefformat{group}{gp.~#2\cirb{#1}#3}
\Crefformat{group}{Group~#2\cirb{#1}#3}
\crefmultiformat{group}{gp.~#2\cirb{#1}#3}{ and~#2\cirb{#1}#3}{, #2\cirb{#1}#3}{ and~#2\cirb{#1}#3}
\Crefmultiformat{group}{Groups~#2\cirb{#1}#3}{ and~#2\cirb{#1}#3}{, #2\cirb{#1}#3}{ and~#2\cirb{#1}#3}
\crefrangeformat{group}{gp.~#2\cirb{#1}#3 to~#2\cirb{#1}#3}
\Crefrangeformat{group}{Groups~#2\cirb{#1}#3 to~#2\cirb{#1}#3}

\newcommand{\mlean}{Lean\xspace} %
\newcommand{\leandy}{LeanDY\xspace}

\newcommand{\mproverif}{\textsc{ProVerif}\xspace}
\newcommand{\msapic}{\textsc{Sapic+}\xspace}
\newcommand{\mStrandsRocq}{\textsc{StrandsRocq}\xspace}

\newcommand{\mtamarin}{\textsc{Tamarin}\xspace}

\newcommand{\mDYstar}{DY${}^*$\xspace}
\newcommand{\mfstar}{F${}^*$\xspace}
\newcommand{\mcomparse}{\textsc{Comparse}\xspace}
\newcommand{\mtulafale}{\textsc{TulaFale}\xspace}
\newcommand{\mrocq}{\textsc{Rocq}\xspace}

\newcommand{\mcheckmate}{\textsc{CheckMate}\xspace}
\newcommand{\mtidy}{\textsc{Tidy}\xspace}
\newcommand{\mwhy}{\textsc{Why3}\xspace}

\newcommand{\append}{\mathbin{{+}\mspace{-4mu}{+}}}

\newcommand{\leandyParenthesizeIfNonempty}[1]{%
  \if\relax\detokenize{#1}\relax\else\left( #1 \right)\fi%
}
\newcommand{\leandySuperscriptIfNonempty}[1]{%
  \if\relax\detokenize{#1}\relax\else^{#1}\fi%
}
\DeclareDocumentCommand{\validExchange}{
  g %
}{%
    \ensuremath{\mathtt{valid\_exchange}%
     \IfValueT{#1}{\leandyParenthesizeIfNonempty{#1}}}%
}
\DeclareDocumentCommand{\validBytes}{
  O{tr} %
  D<>{\Gamma} %
  m
}{%
    \ensuremath{\mathtt{valid\_bytes}_{#1}^{#2}\left( #3 \right)}%
}
\DeclareDocumentCommand{\validState}{
  O{tr} %
  D<>{\Gamma} %
  m
}{%
    \ensuremath{\mathtt{valid\_state}_{#1}^{#2}\left( #3 \right)}%
}
\DeclareDocumentCommand{\validEvent}{
  O{tr} %
  D<>{\Gamma} %
  m
}{%
    \ensuremath{\mathtt{valid\_event}_{#1}^{#2}%
        \left( #3 \right)}
}
\DeclareDocumentCommand{\hasType}{
  O{tr} %
  D<>{\Gamma} %
  m %
  m %
}{%
    \ensuremath{#2\vdash_{#1} #3 : #4}%
}

\DeclareDocumentCommand{\validTrace}{
  o %
  g %
}{%
  \ensuremath{\mathtt{valid\_trace}%
    \IfValueT{#1}{\leandySuperscriptIfNonempty{#1}}%
    \IfValueT{#2}{\leandyParenthesizeIfNonempty{#2}}}%
}

\DeclareDocumentCommand{\mhash}{o}{%
  \IfValueTF{#1}{%
    \ensuremath{\text{hash}_{#1}}%
  }{%
    \ensuremath{\text{hash}}%
  }%
} 
\DeclareDocumentCommand{\msign}{o}{%
  \IfValueTF{#1}{%
    \ensuremath{\text{sign}_{#1}}%
  }{%
    \ensuremath{\text{sign}}%
  }%
} 
\DeclareDocumentCommand{\mverify}{o}{%
  \IfValueTF{#1}{%
    \ensuremath{\text{verify}_{#1}}%
  }{%
    \ensuremath{\text{verify}}%
  }%
} 
\DeclareDocumentCommand{\msenc}{o}{%
  \IfValueTF{#1}{%
    \ensuremath{\text{symenc}_{#1}}%
  }{%
    \ensuremath{\text{symenc}}%
  }%
} 
\newcommand{\mmsenc}[2]{\ensuremath{\left\{ #1\right\}_{#2}^s}}

\DeclareDocumentCommand{\msdec}{o}{%
  \IfValueTF{#1}{%
    \ensuremath{\text{symdec}_{#1}}%
  }{%
    \ensuremath{\text{symdec}}%
  }%
} 
\DeclareDocumentCommand{\maenc}{o}{%
  \IfValueTF{#1}{%
    \ensuremath{\text{asymenc}_{#1}}%
  }{%
    \ensuremath{\text{asymenc}}%
  }%
} 
\DeclareDocumentCommand{\madec}{o}{%
  \IfValueTF{#1}{%
    \ensuremath{\text{asymdec}_{#1}}%
  }{%
    \ensuremath{\text{asymdec}}%
  }%
} 

\newcommand{\mmtuple}[1]{\ensuremath{\left< #1 \right>}}
\DeclareDocumentCommand{\mkLabelAT}{m O{tr}}{
  \left\llbracket #1 \right\rrbracket_{#2}
}

\newcommand{\mBytesEqlst}{\ensuremath{\mathbb{B}}}
\newcommand{\mBytesEq}{\mBytesEqlst\xspace}
\newcommand{\mBytesNFlst}{\ensuremath{\mathbb{B}^{+}}}
\newcommand{\mBytesNF}{\mBytesNFlst\xspace}
\newcommand{\mState}{State\xspace}
\newcommand{\mEvent}{Event\xspace}
\newcommand{\mLabellst}{\ensuremath{\ell}}
\newcommand{\mLabel}{\mLabellst\xspace}
\newcommand{\mLabelAtlst}{\ensuremath{\mathfrak{l}}}
\DeclareDocumentCommand{\mLabelAt}{O{tr}}{
  \ensuremath{\mLabelAtlst_{#1}}\xspace
}
\newcommand{\mLabelLeq}[1]{\ensuremath{\leq_{#1}}}
\newcommand{\mLabelEquiv}[1]{\ensuremath{\cong_{#1}}}

\makeatletter
\let\originferrule\inferrule
\DeclareDocumentCommand \inferrule { s D<>{right} O {} m m }{%
  \IfBooleanTF{#1}%
  {%
    \originferrule*[#2=\textsc{#3}]{#4}{#5}%
  }
  {%
    \originferrule[#3]{#4}{#5}%
  }%
  \IfValueT{#2}%
  {%
    \my@name@inferrule{\textsc{#3}}%
  }%
}
\NewDocumentCommand \my@name@inferrule { m }{%
  \def\@currentlabelname{\ensuremath{#1}}%
}
\makeatother

\newcommand{\kleisli}{\mathrel{\tikz[baseline=-0.5ex]{\draw[{Implies[reversed]}-{Implies[]}, double, double distance=1pt] (0,0) -- (0.3,0);}}}

\AtBeginDocument{%
  }

\begin{document}
% \pagenumbering{arabic}
\setlength{\footskip}{40pt}
\fancyfoot[C]{\thepage}

\title{\leandy: Type-Based and Trace-Based Symbolic Protocol Verification in Lean}

\author{Simon Jeanteur}
\orcid{0009-0005-7298-4883}
\affiliation{%
  \institution{TU Wien}
  \city{Vienna}
  \country{Austria}
}
\email{simon.jeanteur@tuwien.ac.at}

\author{Lorenzo Veronese}
\orcid{0009-0005-0459-6993}
\affiliation{%
  \institution{TU Wien}
  \city{Vienna}
  \country{Austria}
}
\email{lorenzo.veronese@tuwien.ac.at}

\author{Magdalena Solitro}
\affiliation{%
  \institution{Fondazione Bruno Kessler}
  \city{Trento}
  \country{Italy}
}
\email{msolitro@fbk.eu}

\author{Matteo Maffei}
\orcid{0000-0001-8061-1685}
\affiliation{%
  \institution{TU Wien}
  \city{Vienna}
  \country{Austria}
}
\affiliation{%
  \institution{Christian Doppler Laboratory Blockchain Technologies for the Internet of Things}
  \city{Vienna}
  \country{Austria}
}
\email{matteo.maffei@tuwien.ac.at}

\renewcommand{\shortauthors}{Jeanteur et al.}

\begin{abstract}
Computer-aided formal verification is a widely used approach for the symbolic analysis of cryptographic protocols. However, many modern protocols rely on features that remain challenging for existing techniques. In particular, reasoning about state, time-dependent behavior, inductively defined data structures, unbounded executions, and conditional secrecy requires a level of expressiveness that is difficult to reconcile with effective automation. As a result, protocol verification has largely followed two disjoint paths: fully automated methods with limited expressiveness, or interactive proofs in general-purpose theorem provers that offer flexibility but only limited, non-specialized automation.

We present an orthogonal approach that bridges this gap by combining compositional type-based reasoning with trace-based reasoning, enabling modular verification of stateful and unbounded protocols. Guided by the language-and-automation co-design (LAC) principle, our approach delivers protocol-specific automation while retaining high expressiveness.
We implement this framework as the LeanDY library for the Lean proof assistant, building on and extending the design of DY*, and combining protocol-specific automation with interactive proofs. Our framework supports, in a unified setting, a broad class of functional and security requirements, including secrecy and authentication for stateful protocols, as well as recursive conditional secrecy for protocols using XOR.
We formalize SegWit-style blockchain primitives in LeanDY and demonstrate its expressiveness by carrying out an in-depth formalization of payment channels on top of this blockchain model, verifying punishment mechanisms and properties that depend on chain liveness.
\end{abstract}

\maketitle

This is the technical report.

\section{Introduction}
\label{sec:intro}
Over the past decades, computer-aided formal analysis of cryptographic protocols has matured into a widely used set of techniques and has become a key component of real-world protocol analysis.
For example, \mproverif~\cite{proverif} and \mtamarin~\cite{tamarin} have supported the verification of the design of TLS 1.3 \cite{bhargavanProvingTLSHandshake2014} and \textsc{WireGuard} \cite{lippMechanisedCryptographicProof2019,Kobeissi19}; and \mfstar \cite{Swamy11,fstar} has been used to verify the Signal protocol, the core of modern messaging applications \cite{dy-star}.
However, many modern protocols rely on features that remain challenging for existing techniques.
In particular, reasoning about global state, time-dependent behavior, unbounded executions, and \emph{conditional secrecy}, i.e., secrets that must remain private only until a specific condition is met, requires a level of expressiveness that is difficult to reconcile with automation.
These challenges are particularly prominent for blockchain-based protocols, which, despite their recent adoption in applications such as payment-channel networks, have seen fewer machine-aided verification efforts.
The design of such protocols typically relies on a global, high-integrity ledger, requires reasoning across application-level and consensus-level specifications, and often combines safety and liveness arguments.
Additionally, they may use constructions such as XOR, whose algebraic properties make symbolic reasoning notoriously difficult \cite{Kuesters08}.
As a result, protocols such as the Lightning Network remain only partially verified, with the notable exceptions being \cite{Grundmann26}, which analyzed the protocol up to a fixed bound, and \cite{fabianskiFormallyVerifiedLightning2026}, which formalized a simplified abstraction of the protocol.

Symbolic methods for protocol analysis, as formalized by Dolev and Yao \cite{dolev-yao}, can be categorized into two broad areas: trace-based and type-based methodologies.
Trace-based symbolic proofs are based on a global append-only trace of all actions, e.g., sending messages or storing state, performed by protocol participants.
Confidentiality and authentication properties are thus expressed in terms of \emph{reachability} of specific trace configurations (e.g., there is no trace where a secret value is sent to the network).
Type-based proofs, instead, employ types, notably \emph{refinement types} \cite{Bengtson11,Backes14}, to guarantee that well-typed protocol implementations respect some confidentiality, integrity, or authentication invariants (e.g., a program that sends over the network a value with type \emph{high confidentiality} is rejected by the type system).
Both approaches have limitations.
Trace-based verification supports a high degree of automation, as symbolic provers such as \mproverif or \mtamarin can explore all possible attacker behavior and automatically discover concrete attack traces or prove that such traces do not exist.
However, these tools require examining complete protocol executions, which is infeasible for complex protocols composed of multiple modes or those with an unbounded number of rounds.
This degree of automation is well suited to certain classes of protocols, but becomes harder to retain when reasoning compositionally about cross-layer protocols or when mixing safety and liveness requirements.
In contrast, type-based verification offers strong modularity: each function or sub-protocol is type-checked independently, and the type system guarantees that security invariants hold for any (unbounded) composition of well-typed modules.
The main limitation of such advanced type systems is that proofs cannot be fully automated and require manual intervention.

In this paper, we build on the design of \mDYstar~\cite{dy-star}, combining the above approaches into a unified methodology where type-based proofs can be complemented with inductive trace-based reasoning to obtain security proofs expressed at the trace level.
We implement this methodology as the \leandy library for the Lean proof assistant, providing a mechanized framework for symbolic protocol verification with support for XOR and conditional secrecy.
While XOR has been supported in Horn-clause–based approaches \cite{Kuesters08} and conditional secrecy has been studied in refinement-type systems \cite{Backes14}, these features have not previously been combined in a single mechanized framework.

Leveraging Lean's powerful abstractions (e.g., quotients) and the \emph{language and automation co-design} (LAC) \cite{Ricketts14} design principle, we improve on \mDYstar by defining a structured embedded domain-specific language (eDSL) for specifying protocol implementations, which allows for automated type-based reasoning while providing controlled escape hatches (e.g., trace-based reasoning) when more expressivity is required.
Specifically, \leandy offers a smoother transition between automation and interactive proving compared to the previous \mfstar-based approach, enabling users to fall back to the powerful Lean interactive mode when automation is insufficient for protocol-specific goals such as cross-layer reasoning, liveness, or punishment arguments in layer-2 blockchain protocols.

\bfparagraph{Contributions} More concretely, this paper makes the following contributions.
\begin{itemize}
\item We present \leandy, a library for protocol verification in the Lean proof assistant supporting dynamic compromise, mutable state, \emph{conditional secrecy} properties, and XOR.
The library supports a high degree of automation thanks to the type-based approach, while still allowing for interactive proofs in the trace-based model (\autoref{sec:leandy}).
\item We develop and integrate in \leandy a type-based approach to \emph{conditional secrecy} properties for protocols making use of XOR operations, including chained release patterns (\autoref{sec:releasable}).
\item We formalize SegWit-style blockchain primitives in \leandy, providing a mechanized model of Bitcoin-like transaction structure, witness-based scripts, and timelocks (\autoref{sec:lightning}).
\item We show the expressiveness of \leandy by carrying out an in-depth formalization of payment channels on top of this blockchain model, verifying punishment mechanisms and properties that depend on chain liveness (\autoref{sec:lightning}).
\end{itemize}

The complete \leandy library, the formalization of blockchain primitives, payment channels, and additional example protocols are available online~\cite{leandy}.

\section{Labels for Conditional Secrecy of XOR}
\label{sec:releasable}
In this section, we illustrate the security labels supported by \leandy, exemplifying their use when type-checking protocols as well as the associated challenges. As a running example, we consider a blockchain protocol based on hash time-locked contracts (HTLCs) that uses XOR, hashes, and signatures.

\bfparagraph{Terms and Labels}
We define symbolic terms $M$ as follows:
\[
 \begin{array}{ll}
     N & \coloneq x \,\vert\, x \oplus N \\
     C & \coloneq alice | bob | ... \\
     M & \coloneq N \,\vert\, C \,\vert\, H(M) | Sig_M(M) \,\vert\, ... \\
 \end{array}
\]
where $x$ represents an atomic nonce, and $N$ is the XOR ($\oplus$) of multiple nonces; $C$ denotes arbitrary constants (e.g., principal names); $H$ represents hashes, $Sig$ signatures, etc.

We assign each term a unique secrecy label (or type) in the form:
\[
    L \coloneq \mathcal{S}ec \,\vert\, \mathcal{P}ub \,\vert\, \mathcal{P}riv^n \,\vert\, [L_1 / L_2]_E
\]
where we denote public data as $\mathcal{P}ub$, secret data as $\mathcal{S}ec$; we use $\mathcal{P}riv^n$ to represent data that is only readable by the principal $n$ and becomes public after compromise of $n$; and we use the notation $[L_1 / L_2]_E$ for the conditional swap of labels $L_1$ and $L_2$ based on the condition $E$ representing an event of the protocol. For instance, $[{\mathcal{S}ec}/{\mathcal{P}ub}]_{Leak}$ denotes a term that is initially secret and becomes public after the $Leak$ event. Events refer to specific steps of the execution of the protocol and can be used to define its security specification in terms of \emph{correspondence assertions}~\cite{WooLam93}. We assume that events are irreversible, in the sense that once an event is reached during execution, it cannot be undone; consequently, any conditional label swap triggered by such an event is permanent and cannot be reverted.

These confidentiality labels overapproximate the secrecy of symbolic bytes in the following sense: if a term is labeled $\mathcal{P}ub$, it may be known to the adversary. A $\mathcal{S}ec$ value, instead, is guaranteed to never be sent to the attacker. In particular, we require every term that is sent to the network to have the $\mathcal{P}ub$ label, i.e., all data sent to the network is known to the attacker. This requirement allows us to infer the $\mathcal{P}ub$ label for all terms that appear on the network if the protocol and the adversary consistently follow the requirement (i.e., are well typed). Additionally, not every XOR-related term needs an explicit annotation: once labels are assigned to a small set of protocol terms, \leandy derives the labels of algebraically related XOR expressions automatically by taking the meet of the labels of their operands. Note that the labels defined above correspond to the specific instances required for the purposes of this section; however, \leandy supports the definition of arbitrary dynamic labels. We refer to \autoref{sec:leandy:typesystem} for a more formal presentation of the typing rules.

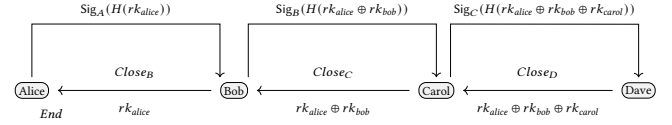
\begin{figure}[t]
\centering
\resizebox{\linewidth}{!}{%
\begin{tikzpicture}[font=\footnotesize]
  \node[draw, rounded corners, fill=black!5, inner sep=2pt] (alice) at (0,0) {Alice};
  \node[draw, rounded corners, fill=black!5, inner sep=2pt] (bob) at (4.0,0) {Bob};
  \node[draw, rounded corners, fill=black!5, inner sep=2pt] (carol) at (8.0,0) {Carol};
  \node[draw, rounded corners, fill=black!5, inner sep=2pt] (dave) at (12.0,0) {Dave};

  \coordinate (abRelL) at ($(alice.east)+(0.18,0)$);
  \coordinate (abRelR) at ($(bob.west)+(-0.18,0)$);
  \draw[<-] (abRelL) -- (abRelR)
    node[midway, below=6pt, align=center, fill=white, inner sep=1pt]
      {$rk_{\mathit{alice}}$}
    node[midway, above=6pt, align=center, fill=white, inner sep=1pt] {$Close_B$};

  \coordinate (asouth) at ($(alice.west)+(0.72,0)$);
  \draw (asouth) node[below=9pt, align=center, fill=white, inner sep=1pt] {$End$};

  \coordinate (bcRelL) at ($(bob.east)+(0.18,0)$);
  \coordinate (bcRelR) at ($(carol.west)+(-0.18,0)$);
  \draw[<-] (bcRelL) -- (bcRelR)
    node[midway, below=6pt, align=center, fill=white, inner sep=1pt]
      {$rk_{\mathit{alice}} \oplus rk_{\mathit{bob}}$}
    node[midway, above=6pt, align=center, fill=white, inner sep=1pt] {$Close_C$};

  \coordinate (cdRelL) at ($(carol.east)+(0.18,0)$);
  \coordinate (cdRelR) at ($(dave.west)+(-0.18,0)$);
  \draw[<-] (cdRelL) -- (cdRelR)
    node[midway, below=6pt, align=center, fill=white, inner sep=1pt]
      {$rk_{\mathit{alice}} \oplus rk_{\mathit{bob}} \oplus rk_{\mathit{carol}}$}
    node[midway, above=6pt, align=center, fill=white, inner sep=1pt] {$Close_D$};

  \coordinate (abTopL) at ($(alice.north)+(0,0.10)$);
  \coordinate (abTopLU) at ($(alice.north)+(0,1.15)$);
  \coordinate (abTopRU) at ($(bob.north)+(-0.28,1.15)$);
  \coordinate (abTopR) at ($(bob.north)+(-0.28,0.10)$);
  \draw[->] (abTopL) -- (abTopLU) -- (abTopRU) -- (abTopR);
  \node[align=center] at ($(abTopLU)!0.5!(abTopRU)+(0,0.28)$)
    {$\mathsf{Sig}_A(H(rk_{\mathit{alice}}))$};

  \coordinate (bcTopL) at ($(bob.north)+(0.28,0.10)$);
  \coordinate (bcTopLU) at ($(bob.north)+(0.28,1.15)$);
  \coordinate (bcTopRU) at ($(carol.north)+(0,1.15)$);
  \coordinate (bcTopR) at ($(carol.north)+(0,0.10)$);
  \draw[->] (bcTopL) -- (bcTopLU) -- (bcTopRU) -- (bcTopR);
  \node[align=center] at ($(bcTopLU)!0.5!(bcTopRU)+(0,0.28)$)
    {$\mathsf{Sig}_B(H(rk_{\mathit{alice}} \oplus rk_{\mathit{bob}}))$};

  \coordinate (cdTopL) at ($(carol.north)+(0.28,0.10)$);
  \coordinate (cdTopLU) at ($(carol.north)+(0.28,1.15)$);
  \coordinate (cdTopRU) at ($(dave.north)+(0,1.15)$);
  \coordinate (cdTopR) at ($(dave.north)+(0,0.10)$);
  \draw[->] (cdTopL) -- (cdTopLU) -- (cdTopRU) -- (cdTopR);
  \node[align=center] at ($(cdTopLU)!0.5!(cdTopRU)+(0,0.28)$)
    {$\mathsf{Sig}_C(H(rk_{\mathit{alice}} \oplus rk_{\mathit{bob}} \oplus rk_{\mathit{carol}}))$};

\end{tikzpicture}%
}
\caption{Conditional release protocol with two intermediaries.}
\label{fig:multihop3}
\end{figure}

\bfparagraph{Proving Correspondence Assertions}
More concretely, let us show how the above labels allow us to prove the correct ordering of protocol events. Consider the protocol depicted in Fig.~\ref{fig:multihop3}. We model a simplified abstraction of the opening and closing operations of payment-channel networks (PCNs) as described in \cite{MSKMR17}. This abstraction describes the conditional release chains present in protocols such as PCNs, atomic swaps using lock contracts, and the Tor network.
Specifically, we assume that a setup phase has been successfully completed and that $bob$ knows the secret $rk_{bob}$ and the assertion $H(rk_{\mathit{alice}} \oplus rk_{\mathit{bob}})$; that $carol$ knows the secret $rk_{\mathit{carol}}$ and the assertion $H(rk_{\mathit{alice}} \oplus rk_{\mathit{bob}} \oplus rk_{\mathit{carol}})$; and that $dave$ knows $rk_{\mathit{alice}} \oplus rk_{\mathit{bob}} \oplus rk_{\mathit{carol}}$. The protocol then proceeds in two phases: first, three channels are opened sequentially between $alice$ and $bob$, $bob$ and $carol$, and $carol$ and $dave$. This is achieved by sending the assertion $Sig(H(rk))$ to the network, where $rk$ represents the revocation secret necessary to close the channel. For the purposes of this example, we abstract away the underlying blockchain mechanics and assume that broadcasting the revocation key to the network suffices to close the channel (see \autoref{sec:lightning} for our encoding of blockchain primitives). To allow for the sequential closing of all intermediate channels, this broadcasting of the revocation keys starts from $dave$, who sends $rk_{\mathit{alice}} \oplus rk_{\mathit{bob}} \oplus rk_{\mathit{carol}}$, allowing $carol$, who knows $rk_{carol}$, to close her channel and broadcast $rk_{\mathit{alice}} \oplus rk_{\mathit{bob}}$. Finally, since $bob$ knows $rk_{bob}$, they can close their channel by broadcasting $rk_{alice}$. We can encode the requirement that all channels are closed (in order) when $alice$ receives $rk_{alice}$ using the following correspondence assertion. %
\[
End \implies Close_B \wedge Close_C \wedge Close_D
\]
This corresponds to the \emph{atomicity} property of payment-channel networks, ensuring that either all payments are completed or none happen.

We prove the above proposition using our conditional confidentiality labels. Let us define an initial labeling as follows.
\[
\begin{array}{lll}
    T_1 &\coloneq rk_{\mathit{alice}} \oplus rk_{\mathit{bob}} \oplus rk_{\mathit{carol}} &: [{\mathcal{S}ec}/{\mathcal{P}ub}]_{Close_D}\\
    T_2 &\coloneq rk_{\mathit{alice}} \oplus rk_{\mathit{bob}} &: [{\mathcal{S}ec}/{\mathcal{P}ub}]_{Close_C \land Close_D} \\
    T_3 &\coloneq rk_{\mathit{bob}} &: [{\mathcal{S}ec}/{\mathcal{P}ub}]_{Close_B} \\
\end{array}
\]
Here, $T_1$, $T_2$, and $T_3$ are identifiers that we can use to refer to the complex XOR expressions. The initial labels match the expected secrecy of the values that are sent on the network: since $T_1$ is the revocation key to close the channel, it is initially secret, but can be published to the network when the event $Close_D$ is reached. Similarly, $T_2$ is initially secret, but can be published only after both $Close_C$ and $Close_D$ have occurred ($Close_C \wedge Close_D$). Finally, in order to showcase our approach for automatically generating labels, we assign an initial label only to $rk_{bob}$, which is considered public after the $Close_B$ event.

We can automatically derive labels for the remaining $rk$ nonces by observing that the XOR of two conditionally confidential terms remains conditionally confidential, with a label given by the conjunction of the events in their respective conditions. Intuitively, the resulting label corresponds to the meet of the initial labels: it is public only if both operands are public, and private otherwise, since the XOR preserves the opacity of the individual nonces.
\[
\begin{array}{lll}
    rk_{\mathit{alice}} &= T_2 \oplus T_3 &: [{\mathcal{S}ec}/{\mathcal{P}ub}]_{Close_B \wedge Close_C \wedge Close_D} \\ 
    rk_{\mathit{carol}} &= T_1 \oplus T_2 &: [{\mathcal{S}ec}/{\mathcal{P}ub}]_{Close_C \wedge Close_D} \\
\end{array}
\]
Since $rk_{alice}$ is $T_2 \oplus T_3$, the condition on the label is the conjunction of the three $Close$ events; similarly, the label of $rk_{carol}$ is the conjunction of the labels of $T_1$ and $T_2$.

Once all values sent to the network have been assigned labels, we proceed as follows. For each message sent, we must establish that its label is public. For each value that is received, we compute the set of events that must have occurred for the value to be sent, under the assumption that its label is public. We analyze each message exchange (input/output pair) performed by each participant in isolation. We begin with $dave$. The value $T_1$ is sent over the network and must therefore be labeled $\mathcal{P}ub$, which is only possible after the occurrence of the $Close_D$ event. Continuing with $carol$, the value $T_2$ is sent and must also be public. 
While $carol$ can locally trigger the $Close_C$ event, establishing that $Close_D$ has occurred requires additional reasoning. In particular, $carol$ can compute $T_2$ only after receiving $T_1$ and combining it with her secret $rk_{carol}$. Since $T_1$ is received from the network, its label must be $\mathcal{P}ub$, which implies that the event $Close_D$ must already have taken place.
A similar argument applies to $bob$. In order to send $rk_{alice}$, $bob$ must execute the $Close_B$ event, and moreover, both $Close_C$ and $Close_D$ must have occurred in prior steps of the protocol. Finally, upon reaching the $End$ event, $alice$ receives $rk_{alice}$ from the network. Since this value must be public, it follows that all three $Close$ events have occurred, thereby establishing our correspondence assertion goal.

The above example highlights the compositional nature of our proof technique. In particular, we establish our security goal by analyzing each protocol participant in isolation, while maintaining the invariant that all data sent over the network is labeled as public. This enables us to derive, for each local action, the set of events that must have occurred to justify the release of a given value. By composing these local arguments, we can obtain a global ordering of events that satisfies the correspondence assertion, thus establishing our goal without requiring a global analysis of the protocol.

\bfparagraph{Handling dynamic compromise} This type-based approach can be extended to handle dynamic compromise of protocol participants. Specifically, we use the $\mathcal{P}riv^n$ label to represent data that is private to principal $n$ but becomes public upon compromise of $n$. We then replace the $\mathcal{S}ec$ label in the initial labeling above with the corresponding $\mathcal{P}riv$ label for each principal. However, $\mathcal{P}riv$ alone is insufficient to model the secrecy of $T_2$ and $T_3$. For instance, when $carol$ is compromised, $T_2$ becomes public only if either $dave$ is also compromised or the $Close_D$ event has occurred. We capture this dependency using the meet operator ($\sqcap$), which yields a label that is public only when both operands are public.
The initial labeling of the values used in the protocol is thus:
\[
\begin{array}{ll}
    T_1 &: [{\mathcal{P}riv^{dave}}/{\mathcal{P}ub}]_{Close_D}\\
    T_2 &: [{\mathcal{P}riv^{carol} \sqcap \mathcal{L}(T_1)}/{\mathcal{P}ub}]_{Close_C \land Close_D} \\
    T_3 &: [{\mathcal{P}riv^{bob} \sqcap \mathcal{L}(T_2)}/{\mathcal{P}ub}]_{Close_B} \\
\end{array}
\]
where we use $\mathcal{L}(T_n)$ to refer to the initial label of $T_n$. Note that the structure of the label matches the chaining of compromises required for the values to be public.

The computed labels follow the same principle as above: the label of an XOR is given by the meet of the labels of its operands.
\[
\footnotesize
\begin{aligned}
r&k_{\mathit{alice}} = T_2 \oplus T_3: \\
& \Big[
(\mathcal{P}riv^{carol} \sqcap \mathcal{L}(T_1))
\sqcap (\mathcal{P}riv^{bob} \sqcap \mathcal{L}(T_2))
\Big/ \mathcal{P}ub
\Big]_{Close_B \wedge Close_C \wedge Close_D}
\end{aligned}
\]

Note that the dynamic compromise of principals must be accounted for in the security goal: we either assume that principals remain uncompromised for the duration of the protocol, yielding the proof above, or we explicitly incorporate the possibility of compromise into the security goal:
\[
  \text{\faIcon{radiation-alt}}_{alice} \,\lor 
  End \implies  \left(\begin{aligned} &(Close_B \lor \text{\faIcon{radiation-alt}}_{bob}) \land (Close_C \lor \text{\faIcon{radiation-alt}}_{carol}) \\ &\qquad \quad\land (Close_D \lor \text{\faIcon{radiation-alt}}_{dave} )\end{aligned}\right)
\]
Here, compromised participants, denoted $\text{\faIcon{radiation-alt}}_{p}$, share their private knowledge with the adversary. Such an adversary can thus broadcast the release secrets without triggering the honest principals' events. This is reflected in the property, where, for each event, either the event is triggered or the respective principal is compromised. The proof follows the steps of the previous example, with an additional case split explicitly handling the possibility of compromise for each participant.

\section{The LeanDY Library}
\label{sec:leandy}
The \leandy library is a \mlean framework for proving \emph{trace properties}, such as authentication and confidentiality, of protocols modeled as state machines that rely on a specific set of APIs, covering cryptographic primitives (encryption, signatures, etc.), network communication, and mutable storage.
Protocols use symbolic cryptography and are verified against a Dolev-Yao-style adversary.
The basic proof technique is to establish a \emph{trace invariant}: every state transition (i.e., any function that modifies the trace, including those modeling the adversary) is proved to preserve this invariant. Consequently, any reachable state, regardless of whether it is generated by an attacker or an honest protocol participant, satisfies the invariant.

Following the design principle of language and automation co-design (LAC)~\cite{Ricketts14}, \leandy provides a general trace invariant, \texttt{valid\_\allowbreak{}trace}, parameterized by protocol-specific behavior, and mechanisms for users to specify only these protocol-dependent components.
In addition, the library offers a Domain-Specific Language (DSL) for protocol specifications and includes a type system that supports the semi-automatic verification, for each function written in the DSL, of the protocol-specific parts of the trace invariant.

More specifically, users of the library write the protocol specifications as a collection of \emph{message exchanges} using the DSL, instantiate the protocol-specific components of the trace invariant (states, events, and validity predicates), and then use the type system together with semi-automated proof support to show that each exchange satisfies $\validExchange{}$.
From these per-exchange proofs, \leandy derives preservation of $\validTrace{}$ for the trace updates generated by the protocol. As a final step, the user provides a separate extrinsic proof that the resulting trace invariant implies the desired security property (e.g., authentication, confidentiality, etc.)

We now present the main building blocks of the \leandy library.
We first describe the core library, which provides the foundational machinery for verification: the definition of traces, the trace invariant, the representation of symbolic bytes, the attacker model, and the type system used to reason about the confidentiality and integrity of messages.
We then turn our attention to the user-facing layer of the library and present the DSL used to specify protocols, and the proof automation included in the library.

\subsection{Core Definitions}\label{sec:leandy:core}

\subsubsection{The Trace and the Trace Invariant}\label{sec:leandy:trace}
The trace represents the global trail of all observable actions performed by protocol participants and adversaries. We implement the trace as a list of \code{Trace-\allowbreak{}Event}s, as reported in the following listing.
\begin{lstlisting}[basicstyle=\small]
inductive TraceEvent (α η : Type) where
| MsgSent (msg : @@Bi α η) : TraceEvent α η
| Corrupt (principal : ℙ) : TraceEvent α η
| SetState (content : @@Bi α η) : TraceEvent α η
| Event (e : Event α η) : TraceEvent α η

abbrev @@T (α η : Type) := List (TraceEvent α η) -- the trace
\end{lstlisting}
The trace records sent messages (\code{MsgSent}); the time point when a principal has been compromised (\code{Corrupt}, which can happen at any time); serialized protocol states (\code{SetState}); and protocol-specific events (\code{Event}) used for, e.g., authentication properties.
Different traces are ordered by suffix, as new events are prepended to the trace. All trace properties in \leandy, e.g., validity or secrecy of messages, are \emph{monotone} with respect to the trace. %
For instance, if a value is public (i.e., accessible to the adversary), it remains so after extending the trace.

We define a trace to be valid, i.e., $\validTrace{tr}$ holds, if, for all events in the trace, each event respects the following property:
\begin{lstlisting}
abbrev valid_trace_event (Γ: LabelingEnv α η) 
        (ev : TraceEvent α η) (tr: @@T α η) : Prop :=
  match ev with
  | MsgSent m => valid_bytes Γ m tr @@wedge m.is_public Γ tr
  | SetState s => match u_state_parse.parse s with
                  | some s =>  u_valid_state Γ s tr
                  | none => False
  | .Event e => u_valid_event Γ e tr
  | .Corrupt _ => True
\end{lstlisting}
Specifically, a message event is valid if the message is public with respect to its confidentiality label and its symbolic bytes are valid (\autoref{sec:leandy:typesystem}). A state is valid if it can be parsed into a valid protocol-specific state, and an event must be valid according to the protocol-specific invariant (\autoref{sec:leandy:state}).

\subsubsection{Symbolic Bytes and the Adversary}\label{sec:leandy:bytes}

Every value sent over the network or stored in a principal state is represented as symbolic bytes. 
Bytes are encoded as the following datatype:

\begin{lstlisting}[basicstyle=\footnotesize]
inductive InnerBytes α η : Type
| Xor (content : Finset η)
| Const (c : Const)
| Key (p : α) (kind : Kind.Key) (k : InnerBytes α η)
| Sign (p : α) (m k : InnerBytes α η)
| Verify (p : α) (σ k : InnerBytes α η)
| Hash (p : α) (m : InnerBytes α η)
| SymEnc (p : α) (m k : InnerBytes α η)
| SymDec (p : α) (c k : InnerBytes α η)
| AEnc (p : α) (m k : InnerBytes α η)
| ADec (p : α) (c k : InnerBytes α η)
| Tuple (p : α) (a b : InnerBytes α η)
| Proj_1 (p : α) (x : InnerBytes α η)
| Proj_2 (p : α) (x : InnerBytes α η)

def Bytes (α η : Type) := Option (InnerBytes α η)
\end{lstlisting}
where we use the \code{Xor} constructor to represent nonces, optionally composed of the bitwise XOR of multiple other nonces; we use the \code{Const} constructor to represent constants, e.g., principal names; the \code{Key} constructor is used to represent keys and contains a label describing its kind (i.e., signing, verifying, or symmetric key) and its bitwise representation (usually a nonce); the rest of the \code{InnerBytes} constructors symbolically represent cryptographic operations (signatures, hashing, symmetric and asymmetric encryption), tuples, and the corresponding destructors.
The $\alpha$ parameter  distinguishes different versions of each cryptographic function, allowing the use in the same protocol, e.g., of different hash functions. 
The set of available nonces, represented by $\eta$, is decided \emph{before} protocol execution.
This is a significant departure from \mDYstar, where nonces are dynamically added to the trace; our approach substantially simplifies the definition of the type system~\autoref{sec:leandy:typesystem}.

The \code{Bytes} type is a nullable \code{InnerBytes}, i.e., we use the special value \code{(none : Bytes)}, also called \code{fail}, to encode failure. For example, applying a projection operator on a value other than a tuple, or verifying an invalid signature returns the \code{fail} value.
This reduction behavior is encoded in the normalization function $\gamma$. The function reduces a \code{Bytes} value to normal form, encoding the symbolic cryptography equational theory: e.g., decrypting an encrypted message with the matching key reduces to the original plaintext.

\bfparagraph{Bytes Quotient and Equational Theories}
To simplify the reasoning about equational theories, \leandy exposes symbolic bytes using Lean's powerful quotient abstraction. Specifically, by considering \emph{bytes modulo equality of their normal form}, we sidestep the problem of explicitly requiring the user to manually reduce constructor, destructor, and equivalence rules by calling $\gamma$.
The main type for symbolic bytes exposed to users is the \code{BytesEq} (or \mBytesEq) quotient type defined as follows.
\[
    \llbracket x \rrbracket : \mBytesEq \coloneq \{ a : \code{Bytes} \mid \gamma(a) = \gamma(x) \}.
\]
That is, we represent symbolic bytes as equivalence classes of bytes that reduce to the same normal form.
This way, when encountering symbolic bytes in protocol code, users of \leandy are not required to handle all possible ways specific bytes can be constructed. Let us consider, for instance, a protocol that decrypts a message, expects a pair, and compares the first projection with a previously stored nonce. Syntactically, \code{tuple(x,y).proj\_1} is built from a projection constructor applied to a tuple constructor, but in \mBytesEq it is equal to \code{x} because both have the same $\gamma$-normal form.
Protocol proofs can therefore use ordinary equality on \mBytesEq as the quotient hides the equational reasoning.

\bfparagraph{Dolev-Yao Attacker}\label{sec:dy-attacker}
We model a Dolev-Yao-style adversary with the inductive predicate \code{AttackerKnows} reported in \autoref{fig:attackerknows}.
\begin{lstlisting}[basicstyle=\scriptsize,float,caption={Definition of \texttt{AttackerKnows}, modeling adversary knowledge.},label={fig:attackerknows}]
inductive AttackerKnows Γ tr : @@Bi α η → Prop
-- public knowledge
| Published {m} (hin : .MsgSent m ∈ tr) : AttackerKnows m
| Fail : AttackerKnows .fail
| Initial {n} (hin : n ∈ Γ.public_names tr) : AttackerKnows (.nonce n)
| Const {c} : AttackerKnows (.const c)
| Zero : AttackerKnows (.xor_set {})
-- constructors and destructors
| Xor : AttackerKnows (.xor_set s₁) → AttackerKnows (.xor_set s₂) → AttackerKnows (.xor_set (s₁ ∆ s₂))
| Key : AttackerKnows k → AttackerKnows (.key p kind k)
| Sign : AttackerKnows m → AttackerKnows k → AttackerKnows (.sign p m k)
| Verify : AttackerKnows σ → AttackerKnows k → AttackerKnows (.verify p σ k)
| Hash : AttackerKnows m → AttackerKnows (.hash p m)
| SymEnc : AttackerKnows m → AttackerKnows k → AttackerKnows (.symenc p m k)
| SymDec : AttackerKnows c → AttackerKnows k → AttackerKnows (.symdec p c k)
| AEnc : AttackerKnows m → AttackerKnows k → AttackerKnows (.asenc p m k)
| ADec : AttackerKnows c → AttackerKnows k → AttackerKnows (.adec p c k)
| Tuple : AttackerKnows a → AttackerKnows b → AttackerKnows (.tuple p a b)
| Proj_1 : AttackerKnows a → AttackerKnows (.proj_1 p a)
| Proj_2 : AttackerKnows a → AttackerKnows (.proj_2 p a)
\end{lstlisting}
The predicate models all information that can be derived by the attacker starting from their initial knowledge. Specifically, the adversary initially knows all public information, including constants (\code{Const}), the \code{fail} value (\code{Fail}), and the nonce composed of all zeros (\code{Zero}); all messages sent to the network (i.e., all messages in the trace, \code{Published}) and all nonces that are marked as initially public for the specific protocol (\code{Initial}). Then, for each constructor and destructor, the adversary can apply them to the bitstrings they know to derive new bitstrings.
These are the traditional attacker rules used by, e.g., \mproverif~\cite{proverif}, that encode, for instance, that the adversary can decrypt encrypted messages if they know the decryption key. The only exception is the rule for XOR: the symbol $\Delta$ denotes the symmetric difference of finite sets of nonce names, reflecting the algebraic properties of XOR.
More generally, the encoding is designed to model a maximally powerful Dolev-Yao adversary. To this end, \code{AttackerKnows} is deliberately unconstrained: any restrictions imposed on honest protocol executions must still permit all terms derivable by the adversary.

\subsubsection{Type System}\label{sec:leandy:typesystem}

\leandy includes a type system to support reasoning about the confidentiality and well-formedness of messages. Specifically, the trace invariant requires that every message $m$ added to the trace, i.e., sent on the network, must be public ($\hasType<\Gamma>{m}{\top}$), and valid ($\validBytes{m}$). We now present our construction of secrecy labels, the typing rules encoding confidentiality, and the definition of $\validBytes{}$.

\bfparagraph{Secrecy Labels}
Secrecy labels in \leandy overapproximate the confidentiality level of bytestrings.
Specifically, if a bytestring is labeled as public, it \emph{may} be shared with the attacker, i.e., the public label is a precondition for sending it to a public channel (as a \code{MsgSent} in the trace).
However, it can remain private.
Conversely, private data is \emph{guaranteed} to never be sent to a public channel or known to the attacker.
We encode labels as general trace properties that express when (i.e., in which traces) the corresponding value may be compromised (\lstinline{corrupt₀}). %
\begin{lstlisting}
structure Label α η where -- latter written $\ell$
  corrupt₀ (tr: Trace α η) : Prop
  monotone: Monotone corrupt₀
\end{lstlisting}
The \texttt{monotone} component of each label contains a proof that its trace property is growing with respect to the suffix ordering of traces. This notion corresponds to the intuition that corruption is permanent: since the attacker never forgets data once public, compromised data will remain compromised for all new traces.

Such an encoding allows us to overcome the limited expressivity of the original \mDYstar labels, which could only represent a small set of compromise conditions, and enables \leandy to reason about conditional secrecy and temporal relationships between events. Moreover, new protocol-specific labels can be defined as part of each protocol's development without requiring modifications to the base library.
For instance, \leandy defined the following standard labels.
\begin{lstlisting}[basicstyle=\small]
def @@Pub := {corrupt₀ _ := True, ...} -- public
def @@Sec := {corrupt₀ _ := False, ...} -- secret
def @@P (p : ℙ) := {corrupt₀ tr := .Corrupt p ∈ tr, ...} -- readable by \texttt{p}
def meet l₁ l₂ := {corrupt₀ tr := l₁.corrupt₀ tr @@wedge l₂.corrupt₀ tr, ...}
def join l₁ l₂ := {corrupt₀ tr := l₁.corrupt₀ tr @@vee l₂.corrupt₀ tr, ...}
\end{lstlisting}
where \lstinline|@@Pub| and \lstinline|@@Sec| denote, respectively, public and private data; \lstinline|@@P| $x$ represents the data readable only by the principal $x$; meet and join combine labels using the $\land$ and $\lor$ of their \lstinline|corrupt₀| conditions.

Labels thus describe \emph{who} can read a given set of messages and \emph{when}.
Following the above definitions of meet and join, we can define a \emph{less public than} ordering of labels, parametrized by the current trace.
\begin{equation*}%
  l_1 \mLabelLeq{tr} l_2 \coloneq \forall tr', tr \leq tr' \to l_1\text{\lstinline|.corrupt₀|}\ tr' \to l_2\text{\lstinline|.corrupt₀|}\ tr'
\end{equation*}
Intuitively, $l_1$ is less public than $l_2$ starting from a specific point in time represented by the trace $tr$ if, for all future traces, when $l_1$ is public then also $l_2$ is public.
When $tr$ is the empty trace, $l_1 \mLabelLeq{\emptyset} l_2$ represents the stronger property that $l_1$ is always less public than $l_2$.
With this definition, we can encode the conditional swap label we used in \autoref{sec:releasable} as follows, where we use the \lstinline|h₀| to ensure monotonicity:
\begin{lstlisting}
def releasable (l₁ l₂: ℓ α η) (e: TraceEvent α η) 
               (h₀: l₁ <=[∅] l₂) : ℓ α η where
      corrupt₀ tr := (if e ∈ tr then l₁ else l₂).corrupt₀ tr
      monotone := ...
\end{lstlisting}
Note that in \leandy we use a more general definition that allows us to use any growing function to conditionally swap two labels, and instantiate \code{releasable} as a special case of the general definition.

The relation $\mLabelLeq{tr}$ captures the information-flow ordering on labels, but for non-empty traces it is only a preorder, i.e., it lacks antisymmetry: two labels may differ syntactically while representing the same confidentiality level from trace $tr$ onward.
Therefore we introduce the corresponding equivalence relation.
\[
  l_1 \mLabelEquiv{tr} l_2 \coloneq l_1 \mLabelLeq{tr} l_2 \wedge l_2 \mLabelLeq{tr} l_1
\]
which encodes which labels are indistinguishable starting from the trace $tr$.
Quotienting labels by this equivalence yields the type $\mLabelAt[tr]$ of labels at trace $tr$:
\[
  \mkLabelAT{l_1}[tr] : \mLabelAt[tr] \coloneq  \left\{ l_2 : \mLabel \mid l_1 \mLabelEquiv{tr} l_2 \right\}
\]
Intuitively, $\mLabelAt[tr]$ forgets historical distinctions that no longer affect confidentiality from trace $tr$ onward.
This quotient significantly simplifies formal reasoning with labels: properties can be stated directly on equivalence classes rather than introducing witnesses for equivalence (e.g., $\exists l, l \mLabelEquiv{tr} \mathcal{P}ub$). More importantly, lifting $\mLabelLeq{tr}$ to the $\mLabelAt[tr]$ equivalence classes yields a complete lattice.
\begin{mprop}[Security Lattice]
  Lifting $\mLabelLeq{tr}$ to \mLabelAt[tr] forms a \emph{complete lattice}. We simply write that order $\leq$.\\
  The equivalence classes of \lstinline|@@Pub|, \lstinline|@@Sec|, \lstinline|meet l₁ l₂| and \lstinline|join l₁ l₂| are respectively the top ($\top$), bottom ($\bot$), meet ($\mkLabelAT{\text{\lstinline|l₁|}} \sqcap \mkLabelAT{\text{\lstinline|l₂|}}$) and join ($\mkLabelAT{\text{\lstinline|l₁|}} \sqcup \mkLabelAT{\text{\lstinline|l₂|}}$) of this lattice.
\end{mprop}

\bfparagraph{Typing Rules for Confidentiality}\label{sec:types-confidentiality}
Proofs of confidentiality that use the above secrecy labels are formulated in terms of the type system shown in \autoref{fig:typesystem}.
The judgment $\hasType[tr]<\Gamma>{b}{l}$ assigns a unique label\footnote{for non-\code{fail} terms} $l \in \mLabelAt[tr]$ to the term $b \in \mBytesEq$ at trace $tr$ depending on its shape.
Specifically, constants, hashes, and the special value \code{fail} are public ($\top$).
Signatures and secret keys preserve the label of their payload. 
The symmetric encryption of a value with a key is public only when the label of the key is \emph{less public} than the label of the message. Similar considerations apply for asymmetric encryption. %
Finally, nonces, represented as the XOR of a finite set of nonces, are labeled according to a labeling environment $\Gamma$, which defines the expected confidentiality properties of the protocol.
$\Gamma$ maps sets of nonces to their initial labels.
It is thus important for this initial assignment to be compatible with the expected behavior of XOR.
\begin{mex}[xor compatibility]\label{ex:xor-compat}
  Let $n_1$, $n_2$ and $n_3$ be nonces.
  If $n_1 \oplus n_2$ and $n_2 \oplus n_3$ are published, then effectively $n_1 \oplus n_3$ is as well.
\end{mex}

\noindent
We generalize \autoref{ex:xor-compat} to finite sets of nonces: if two sets of nonces are public, their XOR combination, i.e., their symmetric difference, must also be public. This yields~\eqref{eq:xor-compat-def}:
\begin{equation}\label{eq:xor-compat-def}
  \Gamma(s_1).\text{\lstinline|corrupt₀|}\ tr \wedge \Gamma(s_2).\text{\lstinline|corrupt₀|}\ tr
    \Rightarrow \Gamma(s_1 \triangle s_2).\text{\lstinline|corrupt₀|}\ tr
\end{equation}
\autoref{eq:xor-compat-def} is then a proof obligation when defining a labeling environment.
We present in \autoref{sec:leandy:autolabel} the automation that \leandy provides for this definition and the automatic derivation of the initial labels we showed in \autoref{sec:releasable}.

\begin{figure}
{\footnotesize
  \begin{mathpar}
    \inferrule[L-Const]{ }{
      \hasType[tr]<\Gamma>{\text{const}(c)}{\top}
    }

    \inferrule[L-Hash]{ }{
      \hasType[tr]<\Gamma>{\text{hash}(n)}{\top}
    }

    \inferrule[L-Fail]{ }{
      \hasType[tr]<\Gamma>{\text{fail}}{\top}
    }

    \inferrule[L-Xor]{
      \mkLabelAT{\Gamma\left( t \right)} = l
    }{
      \hasType[tr]<\Gamma>{\bigoplus\nolimits_{x\in t}x}{l}
    }

    \inferrule[L-Tuple]{
      \hasType[tr]<\Gamma>{a}{l_a} \\
      \hasType[tr]<\Gamma>{b}{l_b}
    }{
      \hasType[tr]<\Gamma>{\text{tuple}(a,b)}{l_a \sqcap l_b}
    }\label{rule:hl:tuple}

    \inferrule[L-Sign]{
      \hasType[tr]<\Gamma>{m}{l}
    }{
      \hasType[tr]<\Gamma>{\text{sign}(m, k)}{l}
    }

    \inferrule[L-PubKey]{
      \textsf{kind} \in \left\{ \textsf{Ver}, \textsf{AEnc} \right\} \\
      \hasType[tr]<\Gamma>{k}{l}
    }{
      \hasType[tr]<\Gamma>{\text{key}(\textsf{kind},k)}{\top}
    }

    \hspace*{-1mm}\inferrule[L-PrivKey]{
      \textsf{kind} \in \left\{ \textsf{Sign}, \textsf{ADec}, \textsf{SEnc} \right\} \\
      \hasType[tr]<\Gamma>{k}{l}
    }{
      \hasType[tr]<\Gamma>{\text{key}(\textsf{kind},k)}{l}
    }

    \inferrule[L-SEnc]{
      \hasType[tr]<\Gamma>{m}{l_m} \\
      \hasType[tr]<\Gamma>{k}{l_k} \\
      l_k \leq l_m
    }{
      \hasType[tr]<\Gamma>{\text{symenc}(m,k)}{\top}
    }\label{rule:c:symenc}

    \inferrule[L-AEnc]{
      \hasType[tr]<\Gamma>{m}{l_m} \\
      \hasType[tr]<\Gamma>{dk}{l_{dk}} \\
      l_{dk} \leq l_m
    }{
      \hasType[tr]<\Gamma>{\text{aenc}(m,\text{key}(\text{aenc},dk))}{\top}
    }\label{rule:c:asymenc}
  \end{mathpar}}
  \caption{Typing rules for confidentiality labels.}
  \label{fig:typesystem}
\end{figure}

\bfparagraph{Well-Formed Symbolic Bytes}
Secrecy labels characterize the confidentiality of symbolic bytes; however, they do not, by themselves, restrict how honest protocol participants may use cryptographic primitives. Therefore, \leandy provides a well-formedness predicate \validBytes{} for symbolic bytes and enforces that every message sent to the network satisfies this predicate.
Users of the library are able to customize the predicate by instantiating the \code{CryptoPreds} type class.
\begin{lstlisting}[basicstyle=\footnotesize]
class CryptoPreds α η where
  u_hash_pred     : LabelingEnv α η → α → (m : @@Bi α η)   → @@T α η → Prop
  u_sign_pred     : LabelingEnv α η → α → (m k : @@Bi α η) → @@T α η → Prop
  u_sym_enc_pred  : LabelingEnv α η → α → (m k : @@Bi α η) → @@T α η → Prop
  u_asym_enc_pred : LabelingEnv α η → α → (m k : @@Bi α η) → @@T α η → Prop
  ... -- proofs of monotonicity
\end{lstlisting}
Each field specifies protocol-specific restrictions on honest cryptographic uses; for example, a protocol may allow signing only messages of a particular shape or only after a specific event has occurred. The class defines default (unrestricted) implementations for all predicates, so for each protocol users can specify only the relevant fields, inheriting the default implementation for the others.

In order to allow for the unconstrained use of cryptography by the attacker, \leandy requires each cryptographic operation to either satisfy the corresponding \code{CryptoPred} or be applied with a public key. For example, symmetric encryption is always permitted when the encryption key is public, and otherwise must satisfy the user-defined predicate \code{u\_sym\_enc\_pred}. Analogous conditions apply to hashing, signing, and asymmetric encryption.
\begin{lstlisting}[basicstyle=\footnotesize]
def hash_pred Γ p m tr :=
  is_public Γ m tr @@vee u_hash_pred Γ p m tr
def sign_pred Γ p m k tr :=
  is_public Γ (key p .Sign k) tr @@vee u_sign_pred Γ p m k tr
def sym_enc_pred Γ p m k tr :=
  is_public Γ (key p .SymEnc k) tr @@vee u_sym_enc_pred Γ p m k tr
def asym_enc_pred Γ p m k tr :=
  is_public Γ (key p .AEncK k) tr @@vee u_asym_enc_pred Γ p m k tr
\end{lstlisting}
The \validBytes{} predicate is defined in terms of the above disjunctions and is reported in \autoref{fig:validbytes}. Constants are always valid, tuples are well-formed whenever each of their components is well-formed, and cryptographic operations must satisfy the corresponding predicates: either the relevant keys are available to the attacker, or the protocol-specific usage predicates must hold.

\begin{figure}
{\footnotesize
  \begin{mathpar}
    \inferrule[V-Fail]{ }{
      \validBytes{\text{fail}}
    }

    \inferrule[V-Xor]{ }{
      \validBytes{\text{xor}(x)}
    }

    \inferrule[V-Const]{ }{
      \validBytes{\text{const}(x)}
    }

    \inferrule[V-Key]{
      \validBytes{k}
    }{
      \validBytes{\text{key}(\textsc{kind},k)}
    }

    \inferrule[V-Tuple]{
      \validBytes{a} \\
      \validBytes{b}
    }{
      \validBytes{\text{tuple}(a,b)}
    }

    \inferrule[V-Hash]{
      \text{\code{hash\_pred}}_{tr}^{\Gamma}(m) \\
      \validBytes{m}
    }{
      \validBytes{\text{hash}(m)}
    }

    \inferrule[V-Sign]{
      \text{\code{sign\_pred}}_{tr}^{\Gamma}(m,k) \\
      \validBytes{m} \\
      \validBytes{k}
    }{
      \validBytes{\text{sign}(m,\text{key}(\text{sign},k))}
    }

    \inferrule[V-SEnc]{
      \text{\code{sym\_enc\_pred}}_{tr}^{\Gamma}(m,k) \\
      \validBytes{m} \\
      \validBytes{k}
    }{
      \validBytes{\text{symenc}(m,\text{key}(\text{symenc},k))}
    }

    \inferrule[V-AEnc]{
      \text{\code{asym\_enc\_pred}}_{tr}^{\Gamma}(m,k) \\
      \validBytes{m} \\
      \validBytes{k}
    }{
      \validBytes{\text{aenc}(m,\text{key}(\text{aenc},k))}
    }
  \end{mathpar}}
  \caption{Core rules defining $\validBytes[]<>{}$.}
  \label{fig:validbytes}
\end{figure}

\bfparagraph{Soundness}
We define the soundness of the \leandy type system in terms of the confidentiality and well-formedness of the data that can be derived by the adversary.

\begin{mprop}[Type-system soundness]\label{prop:type-system}
  For every valid trace, anything known by the attacker is public and well-formed:
  \begin{equation}\label{eq:attacker-type-system}
    \forall b.\, \text{\lstinline|AttackerKnows|}_{tr}^{\Gamma}(b) \to \hasType[tr]<\Gamma>{b}{\top} \land  \validBytes[tr]<\Gamma>{b}.
  \end{equation}
\end{mprop}

\noindent
With this theorem, we ensure that the adversary cannot recover private data starting only from publicly available information and that all adversary operations preserve the well-formedness of public data. Moreover, the theorem shows that we can derive such confidentiality and well-formedness judgments without restricting the capabilities of the attacker.

\subsubsection{Protocol-Specific States and Predicates}\label{sec:leandy:state}
The \validTrace{} invariant refers to protocol-specific components that must be specified as part of each protocol definition. This is achieved using the \code{UserPreds} type class, which extends \code{CryptoPreds} with user events, user states, a validity predicate for states, defined as an instance of the \code{HasValid} type class, and a parser/serializer for states.
\begin{lstlisting}[basicstyle=\footnotesize]
class UserPreds (α η : Type) extends CryptoPreds α η where
  u_valid_event : LabelingEnv α η → Event α η → Trace α η → Prop
  u_valid_event_monotone (Γ e) : Monotone (u_valid_event Γ e)

  u_state : Type
  u_state_has_valid : HasValid α η u_state
  u_state_parse : Parsing.Parsable (@@Bi α η) u_state
\end{lstlisting}
The class defines default implementations for all fields: predicates return \code{True} and monotonicity obligations are proved automatically by simplification, making events trivially valid and imposing no restrictions. This way, \leandy users are required to specify only the fields relevant to their protocol, inheriting the default implementations for the rest. The \code{u\_state\_parse} field connects the symbolic trace, where states are stored as bytes, with the structured state type (\code{u\_state}) used by protocol code.

\subsubsection{Parsing and Serializing Bytes}\label{sec:parsing}

Protocol specifications rely heavily on custom formats to abstract low-level details away.
This abstraction does not always reflect reality, yielding the entire class of \emph{format confusion attacks}~\cite{mavrogiannopoulosCrossprotocolAttackTLS2012,patersonThreeLessonsThreema,wallezTreeSyncAuthenticatedGroup2023}.
There is an ongoing effort to make existing tools aware of such attacks via frameworks like \mtulafale~\cite{tulafale} or, even more relevant to our context, \mcomparse~\cite{comparse}.

The core of \leandy is intentionally built around the symbolic bytes datatype introduced in \autoref{sec:leandy:bytes}. This restriction keeps the implementation simple and limits the soundness arguments of \autoref{sec:leandy:typesystem} to the constructors of the term algebra.
To make this representation practical for protocol specifications, \leandy provides a generic parsing and serialization infrastructure between user-defined data types and symbolic bytes. Protocol implementations can therefore be written in terms of domain-specific types, such as protocol states or structured messages, while the library core and its proofs only need to reason about the underlying symbolic byte algebra.

\bfparagraph{Parsing}
Parsing and serialization between any two types is performed through the \lstinline|Parsable| type class.
\begin{lstlisting}
class Parsable (α β: Type) where
  parse: α → Option β
  serialize: β → α
  sound (x:β): parse (serialize x) = some x
notation "do_parse " x " as " γ => Parsable.parse (β:=γ) x
\end{lstlisting}
where, typically, $\alpha$ is instantiated as \mBytesEq. 
The \lstinline|do_parse| notation is used
as a user-friendly typing annotation for parsing.

Parsing is used throughout \leandy, and all relevant types implement it.
It is also used to abstract away \mBytesEq for state management. %
This genericity enables composing parsing definitions. Notably, \lstinline|Parsable| is a transitive relation between types:
\begin{lstlisting}[basicstyle=\small]
def trans_parsable (γ₁ γ₂ γ₃: Type) 
  (h₁: Parsable γ₁ γ₂) (h₂: Parsable γ₂ γ₃): Parsable γ₁ γ₃
where
  parse := h₁.parse /-*$\kleisli$*-/ h₂.parse -- Kleisli composition~\cite{kleisliEveryStandardConstruction1965}
  serialize := h₁.serialize ∘ h₂.serialize
  sound := ...
\end{lstlisting}
This allows for easily instantiating \lstinline|Parsable| on complex types and is what some of the automation relies on: we derive a trivial parsing to a nested tuple, and then, by transitivity, derive the \lstinline|Parsable| instance all the way to \mBytesEq.

\begin{mrmk}
  We do not enforce bijective parsing.
  This enables support for formats where serialization is loosely specified.
  This can notably matter for blockchain protocols (e.g., \autoref{sec:lightning}), where scripts \emph{can} be written in multiple semantically equivalent ways (e.g., using \code{OP\_NOP}s).
\end{mrmk}

\bfparagraph{Extending the Type System}
We allow the user to extend the notion of \emph{valid public} to more than just \mBytesEq using the \lstinline|HasValidPublic| type class.
Combined with parsing (using the \lstinline|ValidParsable| type class), this effectively corresponds to extending the type system with a new constructor.
\begin{lstlisting}[basicstyle=\footnotesize]
class HasValidPublic (α η γ: Type) where
  is_valid_public (Γ: LabelingEnv α η) (x: γ) (tr: @@T α η): Prop
  is_valid_public_monotone Γ (x: γ): Monotone (is_valid_public Γ x) 

class ValidParsable (α η γ δ: Type) 
  [HasValidPublic α η γ] [HasValidPublic α η δ] [Parsable γ δ]
where
  valid_parse Γ tr (x:γ) (y:δ) (h: is_valid_public Γ x tr) (h': parse x = some y):
      is_valid_public Γ y tr
  valid_serialize Γ tr (y:δ) (h: is_valid_public Γ y tr) :
    is_valid_public Γ (serialize y) tr
\end{lstlisting}
Thus, implementing \lstinline|ValidParsable| effectively adds the following \nameref{rule:valid-serialize} rule to the type system.
\[
  \inferrule[ValidSerialize]{%
    \label{rule:valid-serialize}
    \text{\lstinline|is_valid_public|}_{tr}^\Gamma(x)
  }{
    \validBytes{\text{\lstinline|serialize|}\ x} \\ 
    \hasType{\text{\lstinline|serialize|}\ x}{\top}
  }
\]

\bfparagraph{Automation}\label{sec:parsing:automation}
The \nameref{rule:valid-serialize} rule is part of the type system's automation, and \leandy also features macros to automatically derive \lstinline|Parsable| and \lstinline|ValidParsable| instances.
In particular, all types in \autoref{sec:payment-channels} (states, messages, custom scripts,\dots) have parsing automatically derived.

\subsection{Protocols DSL and Proof Automation}\label{sec:leandy:dsl-lac}
A cryptographic protocol is modeled as a set of atomic actions named \emph{exchanges}.
Each exchange corresponds to a (honest) protocol participant receiving a message from the network, performing a stateful computation on the received bitstring, and sending a new message, optionally updating its state or triggering protocol-specific events.
More specifically, each exchange is a function of the following form.
\[
  \textit{exchange} \coloneq \mathbb{N} \to \mBytesEq \to \mState \to  \mBytesEq \times \mState \times [ \mEvent ].
\]
The input includes a (unique) timestamp, to ensure nonce freshness, a message received from the network, represented as symbolic bytes (\autoref{sec:leandy:bytes}), and a protocol-specific state (\autoref{sec:leandy:state}); the output triple includes a message to be published, a possibly updated state and a list of protocol-specific events.

\begin{mex}\label{ex:running-core}
   Consider the simple authentication protocol~\cite{Busi25}:
\begin{align}
    A \to B &:  \mmtuple{A, B, Na} \\
    B \to A &:  \mmsenc{\mmtuple{Na, A}}{Sk_{A\,B}}
\end{align}
\end{mex}
We model the protocol as three exchanges: \begin{enumerate*}[label=(\roman*)]
\item $A$ sends the tuple $\mmtuple{A, B, Na}$ and stores in its state the three values; 
\item $B$ receives the tuple, computes the symmetric encryption of $\mmtuple{Na, A}$ using the shared key $Sk_{AB}$ and sends it over the network (no state is required);
\item $A$ receives the encrypted bitstring, decrypts it with the shared key, and verifies that both the nonce $Na$ and the identifier $A$ correspond to the ones stored in the previous exchange. 
\end{enumerate*}
\begin{lstlisting}[float,caption={Final exchange of the simple authentication protocol},label={fig:simpleauthcheck}]
def initiator_check : exchange Unit Nonce := 
  λ input state => Id.run do
    let .Init sa sb sna := state | return default
    let dec := input.symdec () (sk sa sb)
    if dec ≠ .fail then
      let na' := dec.proj_1 ()
      let a'  := dec.proj_2 ()
      if sa = a' @@wedge sna = na' then
        return { default with events := [.done sa sb na'] }
    return default
\end{lstlisting}
The last exchange of the protocol is shown in \autoref{fig:simpleauthcheck}.
The type \code{exchange Unit Nonce} represents an exchange of symbolic bytes that uses the concrete \code{Nonce} datatype for nonces (see \autoref{sec:leandy:bytes}).
The exchange is a function that receives an \code{input} and a \code{state} and returns a structure with the three fields described above.
We use a \code{default} structure with an empty list of events, an empty state, and the symbol \code{fail} as the default output in case of error (e.g., when the decryption or equality checks fail).
When the decryption does not \code{fail} and all checks are successful, the exchange returns the event \code{done} representing the principal $A$ confirming the end of the protocol.

Each exchange must be proven correct with an extrinsic proof: the user must provide the associated correctness theorem proving the following proposition.
\begin{multline}\label{eq:valid-exchange-def}
\mspace{-18mu}\validExchange{op} \coloneq \forall (in : \mBytesEq)\, (st : \mState),\\
     \left( \begin{multlined}
      \validBytes{in} 
      \wedge\ \validState{st}
      \wedge\ (\hasType{in}{\top})
    \end{multlined} \right) \to \\
    \left( \begin{multlined}
      \validBytes[tr \append evs]<\Gamma>{msg}
      \wedge\ \validState[tr \append evs]<\Gamma>{st'}\\
      \wedge\ (\forall e \in evs,\, \validEvent[tr]<\Gamma>{e})
      \wedge\ \left( \hasType[tr \append evs]<\Gamma>{msg}{\top} \right)
    \end{multlined} \right)
\end{multline}%
\indent where $op(in, st) = \left\langle msg, st', evs\right\rangle$ \\

\noindent
Specifically, $\validExchange{}$ requires that protocol exchanges preserve the validity of states and messages: given a valid public input and a valid state for a specific trace $tr$ in a labeling environment $\Gamma$, an exchange must ensure
\begin{enumerate*}[label=(\roman*)]
    \item the new message $msg$ is both valid and public for the trace $tr \append evs$,
    \item the newly generated state $st'$ is valid for $tr \append evs$ and
    \item each event in $evs$ is valid with respect to the original trace $tr$.
\end{enumerate*}

Once $\validExchange{}$ has been proved for all exchanges composing a protocol, \leandy automatically provides a proof of
\[\validTrace[\Gamma]{tr} \to \validTrace[\Gamma]{tr'}\] for the trace updates that result from all exchanges. This implies that $\validTrace[\Gamma]{tr}$ holds for all traces generated by the protocol. It is then easy to show, provided the protocol-specific validity definitions are correct, that $\validTrace[\Gamma]{tr}$ implies the security goals of the protocol with a separate extrinsic proof.

\subsubsection{Soundness of \validExchange{}: the Transition Function}\label{sec:leandy:transition}
The above result is achieved by composing the observable effects of exchanges on the global trace using the \code{transition} function:
\begin{lstlisting}[basicstyle=\small]
def transition (op : exchange α η) : Traceful Unit := do
  let input <- Traceful.recv
  let state <- Traceful.getState
  let time  <- Traceful.getTime
  let @@<msg, newState, evs@@> := op time input state
  Traceful.triggerEvents evs
  Traceful.saveState newState
  Traceful.send msg
\end{lstlisting}
For each exchange \code{op}, the transition function fetches a state and a message from the trace and executes \code{op} with the two inputs. The three outputs are added to the trace by first triggering all user events, storing the state, and finally sending the new message. The \code{Traceful} monad implements the functions for trace manipulation, e.g., \code{Traceful.recv} and \code{Traceful.send} for sending and receiving messages, and is implemented as an Option-State-Reader stack. 
\begin{lstlisting}[basicstyle=\small]
structure ChoiceFn where
  msg (tr: @@T α η) : { m: @@Bi α η // (.MsgSent m) ∈ tr }
  state (tr: @@T α η) : { m: @@Bi α η // (.SetState m) ∈ tr }

abbrev Traceful:= ReaderT ChoiceFn (StateT (@@T α η) Option)
\end{lstlisting}
Computations in \code{Traceful} may fail (Option), may perform modifications to the implicit trace (State), and may index into the trace using specific criteria (Reader). In particular, the \code{ChoiceFn} record provides functions to obtain events (a message or a state) from the trace. We use this abstraction to allow for reasoning about an over-approximation of all possible choices for messages or state: by keeping \code{ChoiceFn} universally quantified, we do not make any assumption on how states or messages passed to exchanges are selected. This is critical, as any proof using \code{transition} holds regardless of the choice of messages or states: this simulates the adversarial behavior of sending out-of-order or malformed messages to protocol participants.

The \code{transition} function plays a central role in establishing the main soundness result of \leandy, namely that each valid exchange preserves the validity of the trace with respect to the trace invariant:
\begin{lstlisting}[basicstyle=\small]
theorem transition_valid {ch: ChoiceFn} 
           (op: exchange α η) (tr: @@T α η) 
           (hop : valid_exchange Γ tr op) (h₀: valid_trace Γ tr) 
           (ht : transition op ch tr = some ((),tr')) : valid_trace Γ tr' 
\end{lstlisting}
Given a valid trace \code{tr} and a \validExchange{op}, if the \code{transition} function succeeds (i.e., returns \code{some} value) and returns a new trace \code{tr'}, then \code{tr'} is valid. 
With the \code{transition\_valid} theorem, \leandy provides its users a protocol-specific proof of validity of every protocol trace generated by the valid exchanges that compose the honest implementation of a protocol and all adversarial behavior.

\subsubsection{Exchange-Specific Tactics}\label{sec:leandy:tactics}

By proving trace-level preservation once and for all via \code{transition\_valid}, \leandy shifts the entire verification burden to the static properties of individual message exchanges. Proving these exchange-specific invariants---formulated as \validExchange{}-shaped goals---offers a great starting point for automation.
Recall from \autoref{eq:valid-exchange-def} the shape of \validExchange{}.
It loosely consists of two parts: type-system proofs and user predicate proofs.

\bfparagraph{Type-System Automation}
Type checking is generally a mundane task for computers.
\leandy's type system, however, features arbitrary user-provided predicates (i.e., \lstinline|CryptoPred|s) and a generally undecidable subtyping relation (due to the trace-dependent label ordering in \nameref{rule:c:symenc} and \nameref{rule:c:asymenc}).
Aside from these two aspects, the type system does not feature a general subsumption rule, making type checking mostly deterministic.

\begin{figure}
{\footnotesize
  \begin{subfigure}{\columnwidth}
    \begin{mathpar}
      \inferrule[L-TupleMono]{
        \hasType[tr]<\Gamma>{a}{l} \\
        \hasType[tr]<\Gamma>{b}{l}
      }{
        \hasType[tr]<\Gamma>{\text{tuple}(a,b)}{l}
      }\label{rule:tuplemono}

      \inferrule[L-Proj1]{
        \hasType[tr]<\Gamma>{a}{\top}
      }{
        \hasType[tr]<\Gamma>{\text{proj}_1(a)}{\top}
      }

      \inferrule[L-Proj2]{
        \hasType[tr]<\Gamma>{a}{\top}
      }{
        \hasType[tr]<\Gamma>{\text{proj}_2(a)}{\top}
      }

      \inferrule[L-Verify]{
        \hasType[tr]<\Gamma>{\sigma}{\top}
      }{
        \hasType[tr]<\Gamma>{\text{verify}(\sigma, k)}{\top}
      }
    \end{mathpar}
    \caption{Derived confidentiality labeling rules.}
    \label{fig:destructors-labeling}
  \end{subfigure}

  \vspace{0.8em}
  \begin{subfigure}{\columnwidth}
    \begin{mathpar}
      \inferrule[V-Proj1]{
        \validBytes[tr]<\Gamma>{a}
      }{
        \validBytes[tr]<\Gamma>{\text{proj}_1(a)}
      }

      \inferrule[V-Proj2]{
        \validBytes[tr]<\Gamma>{a}
      }{
        \validBytes[tr]<\Gamma>{\text{proj}_2(a)}
      }

      \inferrule[V-ADec]{
        \validBytes[tr]<\Gamma>{c}
      }{
        \validBytes[tr]<\Gamma>{\text{adec}(c, k)}
      }

      \inferrule[V-SDec]{
        \validBytes[tr]<\Gamma>{c}
      }{
        \validBytes[tr]<\Gamma>{\text{symdec}(c, k)}
      }

      \inferrule[V-Verify]{
        \validBytes[tr]<\Gamma>{\sigma}
      }{
        \validBytes[tr]<\Gamma>{\text{verify}(\sigma, k)}
      }
    \end{mathpar}
    \caption{Derived validity rules.}
    \label{fig:destructors-validity}
  \end{subfigure}}
  \caption{Selected derived rules for destructors and tuples.}
  \label{fig:destructors-typesystem}
\end{figure}

Leveraging this determinism, \leandy provides the \lstinline|prove_*| tactics to prove $\hasType[tr]<\Gamma>{m}{l}$ and/or $\validBytes[tr]<\Gamma>{m}$.
To make such automation usable in practice, we derive the typing rules for destructors, some of which are transcribed in \autoref{fig:destructors-typesystem}.
We can omit some of the expected premises by leveraging the fact that some terms effectively get simplified away; this is notably the case for the \lstinline|valid_bytes| rules.
We specialized the rule \nameref{rule:hl:tuple} into \nameref{rule:tuplemono} to focus the automation on the most common case of public labels.
The decryption rules for labels are not present in \autoref{fig:destructors-typesystem} as they involve more complex quantified formulas.

\bfparagraph{Full Automation of Exchanges}
Beyond simple type-system automation, we build on the extensible automation machinery of \mlean to automate the remaining proof obligations.

To simplify reasoning about control flow, we split execution paths in a manner similar to \mStrandsRocq~\cite{Busi25}.
We then leverage our \lstinline|prove_*| tactics alongside \mlean's built-in simplifier and automated proof search machinery (e.g., the \lstinline|grind| tactic~\cite{demoura_grind}) to dispatch straightforward subgoals.
Ultimately, this leaves only the non-trivial goals that capture the core cryptographic properties of the protocol.

This culminates in the \lstinline|intro_valid_exchange| tactic. We showcase its effectiveness by automatically proving the entire \texttt{valid\_\allowbreak{}exchange} property for the \texttt{initiator\_check} from \autoref{ex:running-core} (defined in \autoref{fig:simpleauthcheck}). The automation gracefully hands control back to the user for the cryptographically relevant case of \lstinline|valid_events|, thereby automating away all security-irrelevant details.
\begin{lstlisting}
theorem valid_initiator_check (tr: @@T Unit Nonce) : valid_exchange.alt Γ tr (initiator_check) := by
  intro_valid_exchange for initiator_check
  case valid_events sa sb sna _ _ => 
    simp_all
    exists sa, sb, sna.1
    grind
\end{lstlisting}

\bfparagraph{\textsc{Aesop}-Based Automation}
\leandy also provides a fine-tuned configuration for the \textsc{Aesop} rule-based tactic~\cite{aesop}.
This enables a backtracking search through the type system rules and helper lemmas as well as basic first-order reasoning.
This approach is significantly more lightweight than the previous one, enabling its use in contexts where the simplifier might otherwise loop (e.g., on extremely large terms or complex user predicates).
Due to its higher degree of automation, our \lstinline|aesop_type_system| tactic can notably make use of the more complex decryption and tuple rules.
In the end, it can single-handedly dispatch the proof for the responder role of \autoref{ex:running-core}:
\begin{lstlisting}
theorem valid_responder (a b : ℙ) (tr: @@T Unit Nonce) :
    valid_exchange Γ tr (responder a b) := by
  aesop_type_system
\end{lstlisting}
Unlike \lstinline|intro_valid_exchange|, however, it has a less graceful failure mode, rarely offering the user meaningful progress on failure.

\subsubsection{Automatic Labeling Environments with XOR}\label{sec:leandy:autolabel}

Manually building a typing environment that follows \autoref{eq:xor-compat-def} may look like a daunting task.
A naive but intuitive method is to consider \autoref{eq:intuitive-xor}: $x_1\oplus x_2$ can only be known by those who know both $x_1$ and $x_2$.
\begin{equation}\label{eq:intuitive-xor}
  \Gamma\left( x_1 \oplus x_2 \right) = \Gamma\left( x_1 \right) \sqcap \Gamma\left( x_2 \right)
\end{equation}
where $\sqcap$ is \autoref{sec:leandy:typesystem}'s \lstinline|meet|.

Then, starting from an assignment $\Gamma_0$ of labels to atomic values (i.e., terms of the form \lstinline[mathescape]|.Xor $\{ n \}$| in our setting), we can naturally and easily build a consistent labeling environment from \autoref{eq:intuitive-xor}.
\begin{lstlisting}
def from_singles (Γ₀: η → ℓ α η) : LabelingEnv α η where
  Λ s := /-*$\bigsqcap_{s_0 \in \texttt{s}} $*-/ Γ₀ /-*$s_0$*-/
  ...
\end{lstlisting}

Looking deeper, it might seem like \lstinline|from_singles| and \autoref{eq:intuitive-xor} are not acceptable.
Already a typical use of the one-time pad breaks them: in that setting $m \oplus k$ is expected to be \lstinline|@@Pub| while \lstinline|from_singles| and \autoref{eq:intuitive-xor} would assign it the label $\Gamma\left( m \right) \sqcap \Gamma\left( k \right)$, meaning $m \oplus k$ could only be published after \emph{both} $m$ and $k$ are published.

Thankfully, we are not constrained to atomic values.
Finite sets form a vector space (over the Booleans, with $\triangle$ as the addition), and \lstinline|from_singles| can be generalized to any basis of that space.
This means the user only needs to focus on defining the labels for the most important terms (the basis), letting \mlean compute the rest.
\begin{lstlisting}[mathescape,basicstyle=\small]
-- b is an $\iota$-indexed family which forms a basis 
def from_basis (b : Basis ι Bool (Finset η)) (Γ₀: ι → ℓ α η): LabelingEnv α η where
  Λ s := $\bigsqcap_{i \in \texttt{(b.repr s)}}$Γ₀ $i$
  ...
\end{lstlisting}
where \lstinline[mathescape]|(b.repr s $i$ : Bool)| is the projection of \lstinline|s| on the $i^{\text{th}}$ vector of \lstinline|b|; as the scalars of this vector space are Booleans, we use set notation to denote the result of this projection.

\begin{mex}[\autoref{sec:releasable}'s labels]
  $\{ rk_{\mathit{bob}} \}$, $\{ rk_{\mathit{alice}}, rk_{\mathit{bob}} \}$, and $\{ rk_{\mathit{alice}},\allowbreak rk_{\mathit{bob}}, rk_{\mathit{carol}} \}$ form a basis of the set of linear combinations of revocation keys.
  In \autoref{sec:releasable}, we assigned them the labels $T_3$, $T_2$, and $T_1$, respectively, and used \lstinline|from_basis| to derive the labels of $\{ rk_{\mathit{alice}} \}$ and $\{ rk_{\mathit{carol}} \}$.
\end{mex}

\section{Verifying Payment Channels}
\label{sec:lightning}
In this section, we present our formalization of payment channels using the \leandy library.
We first describe our encoding of blockchain primitives, which allows us to prove safety properties that depend on chain liveness.
We then present the payment-channel protocol, its encoding within \leandy, and the corresponding safety and liveness proofs.

\begin{mrmk}
  All \mlean \lstinline|structure|s of type \lstinline|Type| in this section derive \lstinline|ValidParsable| (see \autoref{sec:parsing}).
  \mBytesNF denotes the non-failing bytes type (i.e., $\left\{ b : \mBytesEq \middle| b \neq \texttt{fail} \right\}$), and \code{Vec $\tau$ $n$} denotes a list of elements of type $\tau$ of length $n$.
  Moreover, we will use ellipsis (e.g., \enquote{\lstinline|...|}) to omit irrelevant content or inline proofs in \mlean listings.
\end{mrmk}

\subsection{Modeling Blockchain Primitives}

We model a SegWit-style Unspent Transaction Output (UTXO) blockchain inside \leandy.
The model is intentionally close to Bitcoin on the core ledger mechanisms---transaction identifiers, spendable outputs, witness-based scripts, and timelocks---while abstracting the full Bitcoin virtual machine and consensus rules.
This is expressive enough to model real-world blockchain protocols, yet lightweight enough to support mechanized proofs.

The key aspect of the model is that it is built on top of the primitives introduced above via the parsing framework (\autoref{sec:parsing}).
This brings, for free, all the soundness guarantees, composition properties, and proof automation of \leandy to the realm of blockchain protocols.

\subsubsection{Structures}
We first introduce the core datatypes of the model.

\bfparagraph{Pointers}
Bitcoin spends outputs by referring to them through a transaction identifier and an output index.
We model these two ingredients explicitly.
A \code{TXID} is represented as a hash-bearing structure, and a \code{Pointer} packages a \code{TXID} together with a \code{VOUT} index.

\begin{lstlisting}[basicstyle=\small]
structure TXID α η := (hash : @@B α η)
structure VOUT := (n : ℕ)
structure Pointer α η := (txid: TXID α η) (vout: VOUT)
\end{lstlisting}

Crucially, a transaction identifier depends only on the transaction body and not on its witnesses.
As in SegWit, this allows us to compute the \code{TXID} of a transaction before its witnesses are attached, which is essential for breaking spending dependencies, like in channel openings.

\bfparagraph{Transactions}
A transaction carries inputs, witnesses, outputs, and a locktime.
We rule out malformed transactions with mismatched numbers of inputs and witnesses using dependent typing and vectors.
\begin{lstlisting}[basicstyle=\small]
structure Transaction α η where
  (ninputs noutputs : ℕ)
  inputs    : Vec (Input α η) ninputs
  witnesses : Vec (@@B α η) ninputs
  outputs   : Vec (Output α η) noutputs
  locktime  : ℕ
\end{lstlisting}
Each output contains an amount and a locking script \code{scriptPubKey}, while each input contains a pointer to a previous output, an unlocking script \code{scriptSig}, and a sequence number used for relative timelocks.

\noindent
\begin{minipage}{0.48\linewidth}
\begin{lstlisting}[basicstyle=\small]
structure Output α η where
  amount : @@btc
  scriptPubKey : @@B α η
\end{lstlisting}
\end{minipage}
\begin{minipage}{0.48\linewidth}
\begin{lstlisting}[basicstyle=\small]
structure Input α η where
  outptr : Pointer α η
  scriptSig : @@B α η
  sequence : ℕ
\end{lstlisting}
\end{minipage}

\bfparagraph{Blocks and blockchain}
A block is a list of transactions, and a blockchain is a list of blocks.
Rather than treating the chain as a separate global state, we reconstruct it from the protocol trace: whenever a block is published, it is sent as a message event.
The current chain is thus extracted by filtering the trace for block messages.
\begin{lstlisting}[basicstyle=\small]
structure Block α η := (transactions : List (Transaction α η))
structure Blockchain α η := (blocks : List (Block α η))

def List.get_blockchain (tr : Trace α η) : Blockchain α η :=
  List.filterMap
      (λ | .MsgSent msg => do_parse msg as Block α η | _ => none)
      tr.reverse
\end{lstlisting}
This design is crucial for the rest of the development: blockchain facts are treated as ordinary trace facts.
Consequently, the general trace machinery of \leandy applies unchanged: blockchain-related messages live in the same trace as protocol states and events, thereby benefiting from the same monotonicity and validity infrastructure introduced in \autoref{sec:leandy}.
Moreover, this design enforces part of blockchain consensus: effectively, only one canonical blockchain matters.

\subsubsection{Validity}
Validity is defined hierarchically.
Scripts justify individual spends, pointers determine whether a referenced output exists and is still unspent, transactions combine script checks with value preservation, and a blockchain is valid when each of its transactions is valid against its preceding prefix.

\bfparagraph{Scripts}\label{sec:modeling-scripts}
A key feature of Bitcoin-like blockchains is the use of scripts to guard outputs.
Fully implementing the Bitcoin Virtual Machine (VM) in \leandy would significantly increase both proof size and proof engineering overhead.
Instead, we abstract over script semantics with a \code{Script} type class.
\begin{lstlisting}[basicstyle=\small]
structure ScriptInput α η where
  (arg witness stripped : @@B α η)
  (sequence locktime relativeHeight : ℕ)

class Script α η δ where
  eval (self : δ) (input : ScriptInput α η) : Prop
  decidableEval self input : Decidable (eval self input)
\end{lstlisting}
The \code{eval} predicate relates the spent output script (parsed as a value of type $\delta$) to the data available during script execution: the \code{scriptSig}, the witness, the stripped transaction used as the signature preimage, and the fields required for timelock checks.
Requiring \code{eval} to be decidable ensures that scripts remain executable, not just logically specified.

This interface is expressive enough to encode both native scripts and wrappers such as Pay-to-Witness-Script-Hash (P2WSH)~\cite{bip141}. %
In the latter case, the witness reveals both redeeming arguments and the underlying script, whose hash must match the hash stored in the spent output.
\begin{lstlisting}[basicstyle=\small]
structure P2WSH (S : Type) := (hash : @@B α η)
structure P2WSH.Witness := (args script : @@B α η)

instance [Script α η S] : Script α η (P2WSH S) where
  eval self input := Option.getD (dflt := False) $ do
    let {args, script} <- do_parse input.witness as P2WSH.Witness
      | return False
    guard (hash _ script = self.hash)
    return Script.eval (<- do_parse script as S)
      {input with witness := args}
\end{lstlisting}
where we exploit the \lstinline|Option| monad. \lstinline|Option.getD| unwraps the \lstinline|Option| object, returning \lstinline|dflt| on failure.
Thus, instantiating \code{Script} for \code{P2WSH S} amounts to parsing the witness, checking the script hash, parsing the revealed script as an \code{S}, and recursively evaluating it.

\bfparagraph{Pointers}
A pointer is valid if it refers to an existing output of an existing transaction in the current blockchain.
However, payment-channel reasoning requires a stronger condition: an input can only spend an output that has not already been consumed.
For this, we define the set of spendable pointers as the set of all output pointers currently on chain minus those that already occur in transaction inputs.
\begin{lstlisting}[basicstyle=\small]
def spendable_ptrs : Finset (Pointer α η) :=
  (output_ptrs b).toFinset.filter (λ ptr => ptr.txid.unique b)
    \ (input_ptrs b).toFinset
\end{lstlisting}
The additional uniqueness check on \code{TXID}s prevents ambiguity when retrieving an output from its transaction identifier.
Membership in \code{spendable\_ptrs} therefore captures exactly the UTXOs that are available for spending.
This check is slightly stronger than Bitcoin's operational treatment of collisions; our model simply rejects ambiguous cases instead of selecting a specific transaction.
For the setting considered here, this restriction does not impose significant limitations.
Indeed, Bitcoin's \texttt{BIP30}~\cite{bip30} rules out the same ambiguity at the level of unspent outputs: duplicate transaction identifiers are disallowed whenever the earlier transaction still has unspent outputs. Correspondingly, we prove that in every valid chain of our model, all spendable outputs have unique transaction identifiers

\bfparagraph{Transactions}
Transaction validity is parameterized by both a blockchain---intuitively, the chain against which the transaction is checked---and a script language.
A valid transaction must spend only spendable pointers, satisfy the locking scripts of all its inputs, and preserve value.
\begin{lstlisting}[basicstyle=\small]
structure Transaction.valid : Prop where
  unique : (tx.ninputs = 0) → tx.unique b
  spendable : tx.all_input_spendable b
  balance : (tx.ninputs ≠ 0) → tx.valid_balance b
  valid_scripts : tx.valid_eval_scripts b script_t
\end{lstlisting}
The \code{unique} field is relevant only for transactions with no inputs (i.e., coinbase transactions): in that case, there is no spent pointer from which uniqueness could otherwise be recovered, so we require a proof that the resulting \code{TXID} is fresh to match~\cite{bip30}.
For ordinary transactions, \code{spendable} prevents double spending, \code{balance} enforces that outputs do not create money, and \code{valid\_scripts} ensures that every input actually unlocks the output it claims to spend.

A transaction already present on the chain is checked against the blockchain prefix preceding it.
This matches the intended causal reading of blockchain execution: when validating a transaction, only earlier transactions are available as spendable outputs.

\bfparagraph{Blockchain}
Finally, a blockchain is valid when each transaction it contains is valid with respect to its preceding prefix.
In the implementation this is expressed by validating every transaction against its restricted sub-blockchain.
\begin{lstlisting}[basicstyle=\small]
def Blockchain.valid := ∀ tx ∈ b, tx.valid_restr b script_t
\end{lstlisting}
where \code{tx.valid\_restr b} means that \code{tx} is checked not against the full blockchain \code{b}, but only against the sub-blockchain consisting of transactions that precede \code{tx} in \code{b}.
This is the right notion for on-chain validation: when a transaction is appended to the chain, only earlier transactions may provide spendable outputs.

This prefix-based definition is precisely what yields the global safety properties we require.
In particular, it rules out double spends between distinct valid transactions and guarantees that outputs remain stable as the chain grows.

\subsection{Payment Channels}\label{sec:payment-channels}

Payment channels are the core building blocks of payment-channel networks such as Lightning~\cite{ln-main}.
Two parties first lock funds in a joint on-chain output and then perform an unbounded number of off-chain balance updates by exchanging signed replacement transactions.
Only the opening and closing phases require interaction with the blockchain: all intermediate payments remain off chain.

The main difficulty is preventing a party from broadcasting an old commitment transaction that reflects a more favorable balance.
This is addressed with revocation secrets and punishment: each update replaces the previous commitment state and simultaneously reveals the secret needed to punish the owner of that old state if it is ever posted on chain.
Hence, participants do not need to trust the counterparty; they only need the ability to detect a revoked close and publish the corresponding punishment transaction in time.

Our formalization focuses on the core two-party revocation mechanism underlying payment channels.
We model channel opening, an unbounded sequence of off-chain updates, unilateral publication of a commitment transaction, and the corresponding punishment transaction.
We abstract away from features orthogonal to our proof obligations, such as fees and balance negotiation, and cooperative closing.

\begin{mrmk}
This section instantiates the types $\alpha$ and $\eta$ to concrete types, namely \lstinline|Unit| and \lstinline|Nonce| respectively, where:
\begin{lstlisting}
inductive Nonce where
| Rk (a b: ℙ) (t: ℕ) -- revocation key for \texttt{a}
| Sk (a: ℙ) -- secret key (e.g., for signing)
| ...
\end{lstlisting}
\end{mrmk}

\subsubsection{Protocol Overview}
{
\def\hd{0.65cm}%
\def\sp{0.1cm}%
\def\mbm{2.1cm}
\setlength{\fboxsep}{1pt}%

\tikzset{
  msg counter/.style={
    draw, fill=white, font=\bf\scriptsize, circle, inner sep=0.2, minimum width=0.8em
  },
  group counter/.style={
    draw, fill=black, text=white, font=\bf\scriptsize, circle, inner sep=0.2, minimum width=0.8em
  },
  pc diagram/.style={
    node distance=5cm and 1cm,
    font=\sffamily\footnotesize,
    box/.style={rectangle, draw ,text width = 1.9cm, fill=white, drop shadow,
      align=center, minimum height = 0.5cm},
    action/.style={rectangle,text width = 3cm, fill=gray!10,
      align=left, minimum height = 0.5cm},
    mess/.style={font=\sffamily\footnotesize},
    frame/.style={fill=gray!10, draw=black!40}
  }
}

\pgfkeys{
  /pcbox/.is family, /pcbox,
  default/.style={
    top pad=0.6,
    bot pad=0.6
  },
  top pad/.estore in = \pcboxTopPad,
  bot pad/.estore in = \pcboxBotPad,
}

\newcommand{\declareABAlts}{
  \coordinate (Asp) at ($(A)-(-\sp,0)$) ;
  \def\mlastCoord{Asp} 
  \coordinate (Ambm) at ($(A)-(\mbm,0)$) ;
  \coordinate (Bsp) at ($(B)-(\sp,0)$) ;
  \coordinate (Bmbm) at ($(B)-(-\mbm,0)$) ;
}

\DeclareDocumentCommand{\mkarw}{s o m m O{\mlastCoord} m o}{%
  \coordinate (#3) at ($(#5)-(0,#4*\hd)$);

  \def\mlastCoord{#3}%
  
  \path[draw, \IfBooleanTF{#1}{latex-}{-latex}] (#3) 
    to node[%
      \IfNoValueTF{#2}{\IfBooleanTF{#1}{below}{above}}{#2}%
    ,align=center]%
      {\IfValueT{#7}{\refstepcounter{msg}\cir{\themsg}\label{#7}: }#6} 
    (Bsp |- #3);
}

\DeclareDocumentCommand{\mkupdatebox}{O{} m m o d<>}{
  \pgfkeys{/pcbox, default, #1}%

  \path (#2) ++(0, \pcboxTopPad*\hd) coordinate (top_pad);
  \path (#3) ++(0, -\pcboxBotPad*\hd) coordinate (bot_pad);
  \def\mlastCoord{bot_pad}%
  
  \draw[dashed, rounded corners] 
    ($(A |- top_pad)-(\mbm,0)$) rectangle ($(B |- bot_pad)-(-\mbm,0)$);
    
  \IfValueT{#4}{
    \draw ($(B |- top_pad)-(-\mbm,0)$)
      node[left, inner sep=2pt, font=\sffamily\scshape\footnotesize, fill=white] 
      {#4};
  }
  \IfValueT{#5}{
    \draw ($(A |- top_pad)-(\mbm,0)$)
      node[group counter]{%
        \refstepcounter{group}%
        \thegroup\label{#5}%
      };
  }
}

\DeclareDocumentCommand{\mkcheck}{s m m O{\mlastCoord} D<>{checks}}{
  \coordinate (#2) at ($(#4)-(0,#3*\hd)$);
  \def\mlastCoord{#2}%
  
  \IfBooleanTF{#1}{
    \draw (B |- #2) 
      node[right] {#5}
      node {\_};
  }{
    \draw (A |- #2)
      node[left] {#5}
      node {\_};
  }
}

\DeclareDocumentCommand{\mkbar}{m m O{\mlastCoord} m}{
  \coordinate (#1) at ($(#3)-(0,#2*\hd)$);
  \def\mlastCoord{#1}%
  
  \path let \p1 = ($(A)-(\mbm,0)$), \p2 = ($(B)-(-\mbm,0)$), \n1 = {\x2 - \x1} in
    node[box, minimum width=\n1, text width=\n1-0.5cm] 
    at ($(A |- #1)!0.5!(B |- #1)$) {#4};
}

\begin{figure}
  \centering
    \resizebox{\linewidth}{!}{%
      \begin{tikzpicture}[pc diagram]
        \node [box] (A) {Alice};
        \node [box, right of = A] (B) {Bob};

        \declareABAlts

        \mkbar{baropen}{1.2}{Opening}

        \mkarw{cf_req}{1.5}{Common Fund}[msg:set-cf]
        \mkarw*{cf_ack}{0.2}{}
        \mkupdatebox{cf_req}{cf_ack}[Funding]<gp:funding>

        \mkarw{cA0}{1.4}{$\mhash\left( rk_A^0 \right)$}[msg:cA0]
        \mkarw*{cB0}{0.2}{$\mhash\left( rk_B^0 \right)$}[msg:cB0]
        \mkarw*[above]{sigB}{1.8}{$\msign\left( tx_{A\to B}^0, sk_B \right)$}[msg:sigB]
        \mkcheck{chk1}{0.2}
        \mkarw[below]{sigA}{0.2}{$\msign\left( tx_{B\to A}^0, sk_A \right)$}[msg:sigA]
        \mkcheck*{chk2}{0.2}
        \mkupdatebox[top pad=1,bot pad=1]{cA0}{chk2}[Commit $0$]<gp:commit0>

        \mkarw{witA}{1}{$\text{witnesses}_A$}[msg:witA]
        \mkcheck*{chk3}{0.2}
        \mkarw*{witB}{0.2}{$\text{witnesses}_B$}[msg:witB]
        \mkcheck{chk4}{0.2}

        \coordinate (publish) at ($(\mlastCoord)-(0,0.6*\hd)$);
        \def\mlastCoord{publish}%
        \path[draw, double, -latex] (A |- publish) 
          to node[below,align=center]
            {\refstepcounter{msg}\cir{\themsg}\label{msg:publish}: publish common fund to blockchain}
          ($(B |- publish)-(-5mm,0)$);
        
        \mkupdatebox{witA}{publish}[Publish]<gp:publish>

        \mkbar{barupdate}{0.8}{Update Loop}
        \mkarw{cAn}{1.9}{$\mhash\left( rk_A^n \right)$}[msg:cAn]
        \mkarw*{cBn}{0.2}{$\mhash\left( rk_B^n \right)$}[msg:cBn]
        \mkarw*[above]{sigB_upn}{1.8}{$\msign\left( tx_{A\to B}^n, sk_B \right)$}[msg:sigB-upn]
        \mkcheck{upchk1}{0.2}
        \mkarw[below]{sigA_upn}{0.2}{$\msign\left( tx_{B\to A}^n, sk_A \right)$}[msg:sigA-upn]
        \mkupdatebox[top pad=0.8,bot pad=1]{cAn}{sigA_upn}[Commit $n$]<gp:commitn>

        \mkcheck{revkev1}{1.1}<\meventStyle{$\texttt{Released A B \_}$}>
        \mkarw[above]{rvkA}{0.2}{$rk_A^{n-1}$}[msg:rvkA]
        \mkcheck*{revchk1}{0.2}
        \mkcheck*{revkev2}{0.5}<\meventStyle{$\texttt{Released B A \_}$}>
        \mkcheck*{revkevold1}{0.7}<\meventStyle{$\texttt{Old}^B_{n-1}$}>
        \mkarw*[below]{rvkB}{0.4}{$rk_B^{n-1}$}[msg:rvkB]
        \mkcheck{revchk2}{0.2}
        \mkcheck{revkevold2}{0.6}<\meventStyle{$\texttt{Old}^A_{n-1}$}>
        \mkupdatebox[top pad=0.8]{revkev1}{revkevold2}[Revoke $n-1$]<gp:revoken>

        \coordinate (startArrow) at ($(A |- \mlastCoord)-(0.5*\mbm,0.2*\hd)$);
        \coordinate (targetArrow) at ($(B |- barupdate)-(-0.2*\hd,0.8*\hd)$);

        \mkbar{barpunish}{0.8}{Punishment}

        \coordinate (txOnChain) at ($(\mlastCoord)-(0,1.7*\hd)$);
        \def\mlastCoord{txOnChain}%
        \path[draw, double, latex-] (A |- txOnChain) 
          to node[above,align=center]
            {\refstepcounter{msg}\cir{\themsg}\label{msg:txOnChain}: $tx_{B\to A}^k$ on chain}
          ($(B |- txOnChain)-(-5mm,0)$);
        
        \draw ($(A |- txOnChain)-(0,0.6*\hd)$) node[right]{if $rk_B^k$ is in Alice's memory} node{\_};

        \mkcheck{defended}{1}<\meventStyle{\texttt{Defended $rk^k_B$ $tx_p$}}>

        \coordinate (punish) at ($(\mlastCoord)-(0,0.4*\hd)$);
        \def\mlastCoord{punish}%
        \path[draw, double, -latex] (A |- punish) 
          to node[below,align=center]
            {\refstepcounter{msg}\cir{\themsg}\label{msg:punish}: a transaction $tx_p$ that spends $tx_{B\to A}^k$}
          ($(B |- punish)-(-5mm,0)$);
        \mkcheck{postdefended}{0.7}<\meventStyle{\shortstack{\texttt{PostDefended}\\\_ $rk^k_B$ $tx_p$}}>
        
        \mkupdatebox[top pad=0.8,bot pad=0.7]{txOnChain}{postdefended}[Punish]<gp:punish>

        \begin{scope}[on background layer]
          \path[draw, dotted] (A) -- ($(A|-\mlastCoord)-(0,0.2)$);
          \path[draw, dotted] (B) -- ($(B|-\mlastCoord)-(0,0.2)$);

          \draw[->,double,rounded corners] 
          (startArrow) 
          -- ($(B |- startArrow)-(-1*\mbm,0)+(0.2*\hd,0)$) 
          -- ($(B |- targetArrow)-(-1*\mbm,0)+(0.2*\hd,0)$)
          -- (targetArrow);
        \end{scope}
      \end{tikzpicture}%
    }
  \caption{High-level flow of the modeled payment-channel protocol.}
  \label{fig:paymentchannel}
\end{figure}
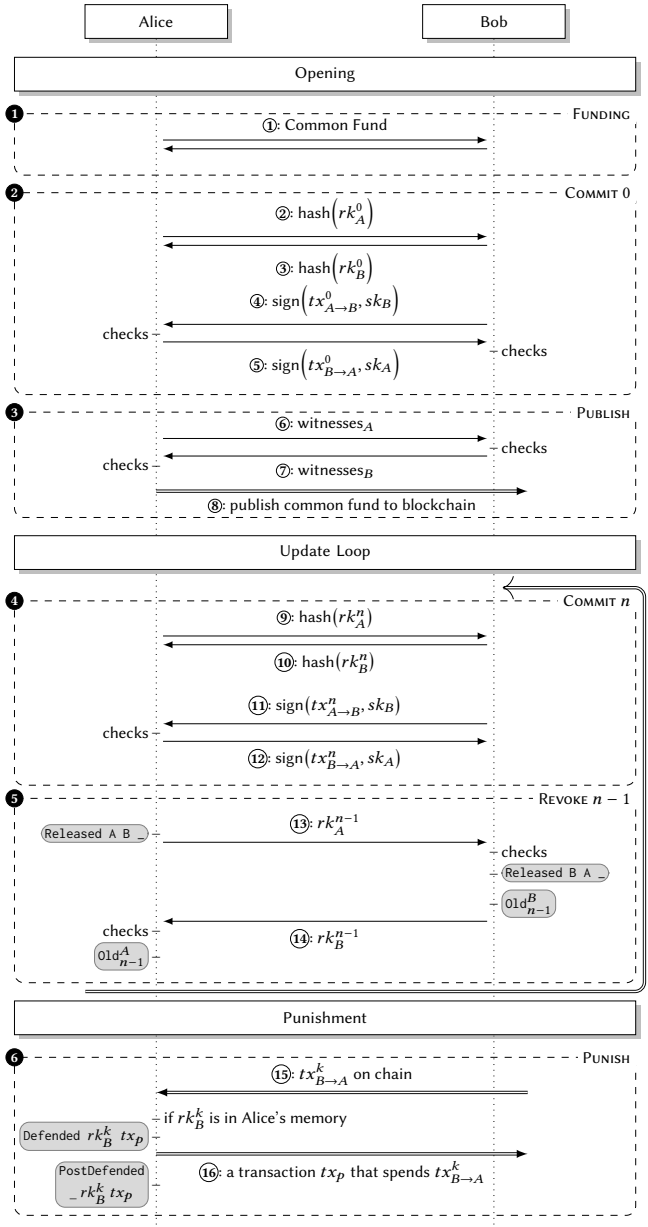
}

The protocol is summarized in \autoref{fig:paymentchannel}.
We model the opening phase, the update phase, and the punishment.
Conceptually, the protocol is built on top of smaller message exchange groups.

\bfparagraph{Funding and Initial Publication}
The opening phase consists of a commitment and two additional groups (\autoref{gp:funding,gp:publish}) to set up the initial common fund used in all subsequent payment-channel transactions.
\autoref{gp:funding} captures the negotiation to agree on the inputs of the common fund.
We bypass modeling this group by universally quantifying over the common fund transaction inputs.
This lets the parties pre-compute the \code{TXID} of the not-yet-published funding transaction.
They wait until \autoref{gp:publish} to exchange the witnesses required to make this transaction valid and publish it, thereby avoiding locking money forever in case \autoref{gp:commit0} fails.

\bfparagraph{Commitments}
When both parties agree on a new balance (via mechanisms beyond the scope of this paper), they exchange enough information for each party to build a unilateral closing transaction.
As modeled in \autoref{gp:commit0,gp:commitn}, these consist of (stripped) transaction signatures and hashes of revocation keys.

\bfparagraph{Revocation}
Both parties can revoke an old channel state by exchanging their respective revocation keys (\autoref{gp:revoken}).
These revocation keys act as proof of good faith that neither party will attempt to publish the now-outdated state on chain.
Failure to uphold this promise results in punishment.
When revocation and commitment are executed sequentially, some of the messages can be merged.
Our model merges \autoref{msg:cBn,msg:sigB-upn} and \autoref{msg:sigA-upn,msg:rvkA}.

\bfparagraph{Punishment}
Finally, \autoref{gp:punish} models the punishment (in \autoref{fig:paymentchannel}, only for Alice).
If a transaction spending the common fund appears on chain and Alice knows its revocation key, Alice sends a transaction that redeems Bob's funds. %
This part of the protocol lives outside of the typical execution flow, as honest agents should never reach it.

\subsubsection{Transactions and Scripts}\label{sec:pc:transaction-and-script}

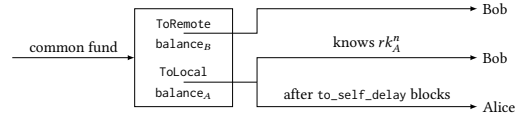
\begin{figure}
  \resizebox{0.8\linewidth}{!}{%
  \begin{tikzpicture}
    \draw[draw] (0,0) rectangle (2,-2);

    \draw[-latex] (-2.5,-1) to node[above] {common fund} (0,-1);

    \draw[-latex] (1, -0.5) node[above]{\code{ToRemote}} node[below]{\footnotesize$\code{balance}_B$} -- (2.5, -0.5) -- (2.5, 0) -- (7, 0) node[right]{Bob};

    \draw (1, -1.5) node[above]{\code{ToLocal}} node[below]{\footnotesize$\code{balance}_A$} -- (2.5, -1.5);
    \draw[-latex] (2.5, -1.5) -- (2.5, -1) -- node[above]{knows $rk_A^n$} (7, -1) node[right]{Bob};
    \draw[-latex] (2.5, -1.5) -- (2.5, -2) -- node[above]{after \lstinline|to_self_delay| blocks} (7, -2) node[right]{Alice};

  \end{tikzpicture}
  }
  \caption{Diagram of the flow of money of $tx_{A\to B}^n$.}
  \label{fig:paymentchannel-tx}
\end{figure}

The security of payment channels relies on their specific choice of transactions and scripts.
The protocol involves two kinds of transactions: 
the \emph{funding transaction}, which has unspecified inputs but is locked by a \code{CommonFund} script; and
the \emph{closing transactions} (e.g., $tx_{A\to B}^n$ in \autoref{fig:paymentchannel}), which spend the aforementioned funding transaction and whose outputs are locked by \code{ToRemote} and \code{ToLocal} scripts.
A schematic of the money flow in a closing transaction is provided in \autoref{fig:paymentchannel-tx}.

Furthermore, payment channels and our model use SegWit-style Pay-to-Witness-Script-Hash (P2WSH) scripts like those introduced in \autoref{sec:modeling-scripts}.
Overall, the scripts under consideration are of type \code{P2WSH Script}, with \lstinline|Script| structured as follows:
\begin{lstlisting}[basicstyle=\small]
inductive Script where
| CommonFundS  (s : CommonFund)  -- funding output
| ToRemoteS    (s : ToRemote)     -- remote commitment output
| ToLocalS     (s : ToLocal)      -- delayed / punishable output
| ...
\end{lstlisting}
We explore each of these scripts, showcasing
the ability of \leandy to capture a wide variety of Bitcoin scripts.

\bfparagraph{\texttt{CommonFund}} %
The common fund transaction is guarded by a standard 2-of-2 multisig script~\cite{bip11}. %
It can be modeled as follows:

\begin{lstlisting}[basicstyle=\footnotesize]
structure CommonFund := (pka pkb : @@B Unit Nonce)
structure CommonFund.Redeem := (signa signb : @@B Unit Nonce)

open CommonFund in instance : Script CommonFund where
  eval self input := Option.getD (dflt:=False) $ do
    let {signa, signb} <- do_parse input.witness as Redeem
    return  verify () signa self.pka = input.stripped @@wedge
            verify () signb self.pkb = input.stripped
  decidableEval := ...
\end{lstlisting}

\bfparagraph{\texttt{ToRemote}}
The \emph{remote} party of a closing transaction is the party that does not store this transaction in their memory (e.g., Bob for $tx_{A\to B}^n$).
This party can retrieve their funds unconditionally and immediately from the closing transaction.

\begin{lstlisting}[basicstyle=\footnotesize]
structure ToRemote := (pk : @@B Unit Nonce)
structure ToRemote.Redeem := (sign : @@B Unit Nonce)

open ToRemote in instance : Script ToRemote where
  eval self input := Option.getD (dflt:=False) $ do
    let m <- do_parse input.witness as Redeem
    return verify () m.sign self.pk = input.stripped @@wedge ...
  decidableEval := ...
\end{lstlisting}

\bfparagraph{\texttt{ToLocal}}
Much of the power of the payment channel comes from the \code{ToLocal} output.
This is the output a party is expected to spend to retrieve their share of the balance when closing unilaterally.

Since unilateral closing can occur without consulting the counterparty, the protocol must ensure that the channel is not closed in an outdated state.
This is where the punishment mechanism is involved.
This output can be redeemed by either party: the closing party must wait for \code{to\_self\_delay}, which allows the counterparty time to execute the punishment.

\begin{lstlisting}[basicstyle=\footnotesize]
structure ToLocal.PunishRedeem := (sign rk : @@B Unit Nonce)
structure ToLocal.RegularRedeem := (sign : @@B Unit Nonce)

open ToLocal in
inductive ToLocal.Eval : ToLocal -> ScriptInput Unit Nonce -> Prop
| regular
    (hp: parse input.witness = some (RegularRedeem.mk sign))
    (hv: verify () sign self.pk_regular = input.stripped)
    (htime: self.to_self_delay ≤ input.sequence @@wedge ... ): -- needs to wait
      ToLocal.Eval self input
| punish
    (hp: parse input.witness = some (PunishRedeem.mk sign rk))
    (hv: verify () sign self.pk_punish = input.stripped)
    (hrk: hash () rk = self.hrk_punish):
      ToLocal.Eval self input

instance : Script ToLocal where
  eval := ToLocal.Eval
  decidableEval := ...
\end{lstlisting}
In other words, \code{ToLocal} offers two redemption paths.
The regular owner can recover the output with a valid signature only after waiting for \code{to\_self\_delay} (the \lstinline|regular| case), while the counterparty can spend it immediately by providing a valid punishment signature together with a revocation key whose hash matches \code{hrk\_punish} (the \lstinline|punish| case).
This is the core revocation mechanism of Lightning-style channels.

\subsubsection{States, Labels, Events and their Invariants}\ \par
\bfparagraph{State}
Each party continuously keeps track of the current channel state and (if applicable) the previous one in a \lstinline{structure} called \code{PcState}:
\begin{lstlisting}[basicstyle=\footnotesize]
structure PcState where
  (self_rk: @@B Unit Nonce)
  (other_rk other_hrk valid_sign : Option (@@B Unit Nonce))
  ...
\end{lstlisting}
where \code{self\_rk}, \code{other\_rk}, \code{other\_hrk}, and \code{valid\_sign} are respectively the party's revocation key, the other party's revocation key and revocation hash, and the local transaction's signature by the other party.

The state invariant ensures consistency among all elements; notably, that \code{valid\_sign} is a valid signature of the correct transaction, and that \code{other\_rk} is the preimage of \code{other\_hrk}.

\bfparagraph{Labels}
This protocol features a comparatively simple set of labels, as the variables are either long-term secrets (e.g., keys with private labels) or public.
The only exception is the revocation key, for which we directly use the conditional-swap operator defined in \autoref{sec:leandy:typesystem}:
\begin{equation*}
  \Gamma\left( \code{Rk a b i} \right)\coloneq
  \text{\lstinline[mathescape]|releasable (@@P a) @@Pub $\ \meventStyle{\texttt{Released a b i}}\ $|} ...
\end{equation*}
Thus, a revocation key is initially secret and becomes public exactly when the corresponding \meventStyle{\texttt{Released \_ \_ \_}} event is emitted at the revoke step of \autoref{fig:paymentchannel}.

\bfparagraph{Events}
Besides the simple events marking the successful execution of an exchange (which are omitted from \autoref{fig:paymentchannel}), our model features the \meventStyle{$\mathtt{Old}^k_I$} events.
These indicate that a party considers a channel state to be outdated, and must therefore defend against it.
These \meventStyle{$\mathtt{Old}^k_I$} events are the triggers for liveness queries.
The \meventStyle{\texttt{Defended rk tx}} and \meventStyle{\texttt{PostDefended \_ rk tx}} events mark successful punishment using a revocation key \texttt{rk} and a punishing transaction \texttt{tx}.
This last event can further only be triggered with a transaction \texttt{tx} which spends a transaction that spends the common fund, capturing that it is a punishing transaction.

\bfparagraph{Signing Invariant}
The most important invariant in our model is imposed on signatures via the \code{CryptoPreds} type class introduced in \autoref{sec:leandy:typesystem}, which constrains honest agents to only sign valid transactions\footnote{at least with the keys reserved for this protocol}.
Furthermore, these transactions are constrained to be either regular payment-channel transactions (e.g., $tx_{A\to B}^n$) or punishment transactions.
\begin{lstlisting}[basicstyle=\footnotesize]
instance instCryptoPreds : CryptoPreds Unit Nonce where
  u_sign_pred Γ _ m k tr :=
    regular_tx_pred tr m k @@vee punishment_tx_pred tr m k
  ...

def regular_tx_pred _ m k := ∃ ..., 
  m = serialize ((mk_tx vks tx_owner cf_ptr _ hrk).strip b 0 ...) @@wedge
  key () .Verify k = vks[signer] @@wedge
  cf_ptr.points_to_common_fund vks b @@wedge /-* \label{lst:invariant:points_to_common_fund} *-/
  ...

def punishment_tx_pred tr stx k := 
    /-*\meventStyle{\texttt{Defended \_ stx}}*-/ ∈ tr @@wedge
    ... -- stx is a stripped punishment transaction
\end{lstlisting}
Here, \code{mk\_tx} builds a closing transaction as per \autoref{sec:pc:transaction-and-script} (e.g., \enquote{\lstinline[mathescape]|mk_tx ($A$, $B$) $A$ cf_ptr _ (hash _ $rk_B^n$)|} is $tx_{A\to B}^n$) that spends \lstinline|cf_ptr|.
\autoref{lst:invariant:points_to_common_fund} further enforces \lstinline|cf_ptr| to be the funding transaction.
Finally, \lstinline|strip| prepares the transaction for signature as per the Bitcoin specification~\cite{bip144}.

\subsubsection{Safety}
We show that the protocol upholds various correctness and safety properties regarding its transactions and the local state of the parties.

\bfparagraph{Correctness of Transactions}
Our model enables us to prove key lemmas about the transactions.
Notably, we can guarantee that all transactions involved have the correct structure.

\begin{mlemma}[Transactions]
  Barring key compromises, a funding transaction can only be spent by a transaction built with \lstinline|mk_tx| using the correct parameters; and
  a \lstinline|ToLocal| output can only be spent by a punishment transaction produced in \autoref{gp:punish}.
\end{mlemma}

\bfparagraph{Correctness of Stored Transactions}\label{sec:correctness-of-stored-tx}
We consider a channel state to be \emph{ready} when a signature and a revocation key hash are known.
\begin{mlemma}\label{lem:ready-state-imp_valid-tx}
  After the check steps in \autoref{gp:commit0,gp:commitn}, each party can construct a valid transaction that spends the common funds.
\end{mlemma}

\noindent
Formally, we prove:
\begin{lstlisting}[basicstyle=\footnotesize]
def isReady : PcState → Prop
| {other_hrk:= some _, valid_sign_stripped := some _, ..} => True
| _ => False

theorem ready_state_imp_valid_tx {tr t pstate pc} -- i.e., \autoref{lem:ready-state-imp_valid-tx}
  (hv : valid_trace Γ tr) 
  (hstate : tr.IsStateAt t (State.Std pstate)) -- pstate is in the trace
  (hmem : pc ∈ state) (hready : pc.isReady) -- a channel state is in pstate
  (hnot_spent : pstate.common_fund ∉ tr.get_blockchain.input_ptrs) : -- the common fund has not been spent
  (pstate.mk_transaction tr.get_blockchain pc).valid tr.get_blockchain (P2WSH Script) := ...
\end{lstlisting}
where \lstinline|pstate.mk_transaction tr.get_blockchain pc| constructs the closing transaction.

\subsubsection{Liveness}
The security of payment channels relies on liveness.
Indeed, the punishment mechanism only functions if the participants are sufficiently reactive to execute the punishment.
Unfortunately, such properties are typically out of reach of standard Dolev-Yao models.
Indeed, such models assume an adversary with full control over the network.
This allows the adversary to permanently block a party from interacting with the protocol, which would prevent the punishment of a cheating counterparty.
Thus, traditional Dolev-Yao-based verification tools such as \mproverif~\cite{proverif} and \mtamarin~\cite{tamarin} do not natively support such properties.
In contrast, \leandy provides the flexibility to reason directly in \mlean about arbitrary trace properties: to \mlean, traces are simply lists. %

Payment channels are safe under the liveness of two entities: the defending party (ensuring that the punishment transaction is sent) and the miners (ensuring that the transaction is eventually included in the blockchain). We first formalize these two notions, then we present our main security theorem.

\bfparagraph{Participant Liveness}
We first define a mechanism to \emph{select} an \lstinline|exchange| for execution.
The core component is a predicate \lstinline|P| that, given a trace, selects a specific \lstinline|exchange| along with its associated message and state.
We define this selection mechanism as follows:
\begin{lstlisting}[basicstyle=\footnotesize]
structure Liveness.Select [up : UserPreds α η]
  (P : @@T α η → exchange α η → @@t → @@t → Prop) (op : exchange α η)
  (tr) (t_msg t_s: @@t) (msg:@@Bi α η) (s: up.u_state) : Prop where
  hop_select: P tr op t_msg t_s
  hmsg: tr.IsMessageAt t_msg msg
  hs: tr.IsStateAt t_s s
\end{lstlisting}
Liveness then guarantees that whatever exchange is selected by \lstinline|Liveness.Select| (if any) is eventually executed, provided the trace grows by a sufficient number of blocks (\lstinline|n|).
\begin{lstlisting}[basicstyle=\footnotesize]
def Liveness [up : UserPreds α η]
  (P : @@T α η → exchange α η → @@t → @@t → Prop) (n : ℕ)
  (tr tr'): Prop :=
  ∀ (t_msg t_s: @@t) (msg:@@Bi α η) (s: up.u_state) op,
    (hselect : Liveness.Select P op tr t_msg t_s msg s) → -- \texttt{op} is selected
    (hgrow: tr <= tr') →
    (hheight: tr.get_blockchain.height + n ≤ tr'.get_blockchain.height) → -- the traces are related and sufficiently many blocks have been mined
    ∃ l₁ l₂, 
      tr' = l₂ ++ (op (l₁.length + tr.length) msg s).as_list ++ l₁ ++ tr -- then \texttt{op} is executed somewhere between \texttt{tr} and \texttt{tr'}
\end{lstlisting}

\bfparagraph{Miner Liveness}
Blockchain consensus algorithms rely on the presence of a sufficient number of honest miners.
Consequently, miners will eventually add all valid transactions to the blockchain.

This validity constraint is captured by \lstinline|ValidLongEnough|, which ensures that a transaction remains valid for at least \lstinline|n| blocks after a given timepoint \lstinline|t|.
\begin{lstlisting}[basicstyle=\footnotesize]
  abbrev Transaction.ValidLongEnough (n : ℕ) (tr) (t:@@t) (tx) : Prop :=
  ∃ tr₁, 
    tr₁ ≤ tr @@wedge 
    (tr.get_blockchain_at t).height + n ≤ tr₁.get_blockchain.height @@wedge
    tr₁.get_blockchain.height < tr.get_blockchain.height @@wedge 
    tx.valid tr₁.get_blockchain script_t
\end{lstlisting}
where \lstinline|get_blockchain_at| retrieves the blockchain state from the prefix of \lstinline|tr| up to time \lstinline|t|.

Assuming miner liveness then means that if a transaction is broadcast to the miners (captured by the \lstinline|published_tx_at| predicate) and remains valid for at least \lstinline|n| blocks, it will be included in the blockchain within those \lstinline|n| blocks.
\begin{lstlisting}[basicstyle=\footnotesize]
def miner_liveness (n : ℕ) tr (t:@@t) tx : Prop :=
  (h₁: published_tx_at tr tx t) →
  (hv: tx.ValidLongEnough script_t n tr t) →
    tx ∈ tr.get_blockchain
\end{lstlisting}

\bfparagraph{Liveness of Punishment}
\begin{lstlisting}[basicstyle=\footnotesize,float,caption={Liveness of Punishment}, label={lst:full-liveness}]
inductive DefenseSelector tr : ... → Prop where
  | intro {...} (hold: .Old {i, ...} ∈ tr) ...: DefenseSelector tr (/-*\autoref{gp:punish}*-/) ... i

theorem punish.eventually_on_chain
  (tr tr_final) (ps) (common_fund) (punish_time miner_time : ℕ)
  (hgrow : tr ≤ tr_final)/-*\label{lst:full-liveness:hgrow}*-/
  (hheight_final : tr.get_blockchain.height + param.toSelfDelay ≤ tr_final.get_blockchain.height)/-*\label{lst:full-liveness:hheight-final}*-/
  (htime_budget : punish_time + miner_time < param.toSelfDelay)/-*\label{lst:full-liveness:htime-budget}*-/

  (hv₁: valid_trace Γ tr_final) (hv₂: tr_final.get_blockchain.valid (P2WSH Script))/-*\label{lst:full-liveness:hv1}\label{lst:full-liveness:hv2}*-/
  (hnot_corrupt : Corrupt ps.a ∉ tr_final @@wedge Corrupt ps.b ∉ tr_final)/-*\label{lst:full-liveness:hnot-corrupt}*-/

  (hcf : common_fund.points_to_common_fund ps.vk tr.get_blockchain ...)/-*\label{lst:full-liveness:hcf}*-/
  (hspent: common_fund.spent tr.get_blockchain)/-*\label{lst:full-liveness:hspent}*-/
  (h_old: ∀ tr' hrk self balance, tr' ≤ tr_final →/-*\label{lst:full-liveness:h-old}*-/
    common_fund.spent_by_hrk (tr'.get_blockchain) ps ... self hrk balance →
    ∃ rk i, hrk = .hash _ rk @@wedge (.Old {ps, self, common_fund, i, rk, balance} ∈ tr')) 

  -- punishment liveness
  (hlpunish : ∀ tr tr_punish,/-*\label{lst:full-liveness:hl-punish}*-/ tr ≤ tr_final → tr_punish ≤ tr_final → 
    Liveness DefenseSelector punish_time tr tr_punish)
  -- miner liveness
  (hlminer : ∀ t tx,/-*\label{lst:full-liveness:hminer}*-/
    miner_liveness (P2WSH Script) published_to_miners miner_time tr_final t tx)

  (hno_race : ∀ rk stx stx' tr', tr' ≤ tr_final →/-*\label{lst:full-liveness:hno-collision}*-/
   (.Defended rk stx ∈ tr') → (.Defended rk stx' ∈ tr') → stx' = stx):
  ∃ rk tx tx',/-*\label{lst:full-liveness:query-start}*-/
    (.PostDefended common_fund rk tx ∈ tr_final) @@wedge/-*\label{lst:full-liveness:query-post-defended}*-/
    tx' ∈ tr_final.get_blockchain @@wedge
    tx.strip tr_final.get_blockchain 0 ... = tx'.strip tr_final.get_blockchain 0 .../-*\label{lst:full-liveness:query-end}*-/
  := ...
\end{lstlisting}
\autoref{lst:full-liveness} features the exact theorem proven to express punishment liveness.
Intuitively, assuming the liveness of both the participants (\autoref{gp:punish}) and the miners, we show that if a revoked commitment transaction is published on the blockchain, then a punishing transaction is accepted on chain within \lstinline|punish_time + miner_time| blocks (representing the response time of the participant and the miners, respectively).
However, due to the highly composable nature of \leandy protocols and the reality of Bitcoin transaction signing, further refinements were required.
We detail these refinements below.

The theorem relates a trace \lstinline|tr| (which contains a revoked commitment transaction) to its extension \lstinline|tr_final|, which we prove contains the corresponding punishment transaction.
Hypotheses on lines \ref{lst:full-liveness:hgrow} to \ref{lst:full-liveness:htime-budget} relate these two traces: \lstinline|tr_final| is a sufficiently long extension of \lstinline|tr| ensuring that enough blocks have been mined to allow the punishment transaction to be sent (\lstinline|punish_time|) and accepted by the miners (\lstinline|miner_time|).
These two liveness delays must fit within the script's \lstinline|toSelfDelay| to ensure that the punishment happens before the cheater can claim the funds.
Hypotheses \lstinline|hv₁| and \lstinline|hv₂| (\autoref{lst:full-liveness:hv1}) enforce generic validity assumptions on the traces.
As shown in \autoref{sec:leandy:transition}, trace validity is an invariant preserved under honest protocol execution; hence, assuming the validity of the extended trace is sound, as honest actions do not violate the transition rules.
\lstinline|hnot_corrupt| (\autoref{lst:full-liveness:hnot-corrupt}) enforces honest signing.
This is also acceptable, as cheating is allowed behavior (it will simply result in punishment) even for honest agents, as per \autoref{sec:correctness-of-stored-tx}.
Hypotheses \lstinline|hcf| and \lstinline|hspent| (\autoref{lst:full-liveness:hcf,lst:full-liveness:hspent}) bind the \lstinline|common_fund| pointer to the common fund of the channel and assert that it has been spent.
\lstinline|h_old| (\autoref{lst:full-liveness:h-old}) further ensures that the transaction spending the common fund corresponds to a revoked state (i.e., the relevant \meventStyle{$\texttt{Old}$} event occurred).
Liveness of \autoref{gp:punish} and of the miners is specified in \autoref{lst:full-liveness:hl-punish,lst:full-liveness:hminer}, respectively.
\lstinline|hlpunish| is universally quantified over subtraces of \lstinline|tr_final| because the defending party must react regardless of the exact block height at which the cheating transaction was published.
Finally, \autoref{lst:full-liveness:hno-collision}'s \lstinline|hno_race| ensures that effectively only one punishment thread succeeds at a time by requiring punishment to use a similar transaction.
It is impossible and meaningless to enforce strict equality on the transaction due to the malleability of SegWit transaction signatures, since witnesses are not signed~\cite{bip143}. %

The theorem's conclusion (Lines \ref{lst:full-liveness:query-start} to \ref{lst:full-liveness:query-end}) must similarly take malleability into account.
\autoref{lst:full-liveness:query-post-defended} states that \lstinline|tx| is a punishment transaction broadcast to the miners.
Since we cannot guarantee it will reach the blockchain completely unaltered due to malleability, we instead prove that the stripped transaction (which excludes witness data) matches the transaction eventually included on-chain.

\section{Related Work}
\label{sec:relatedwork}
\bfparagraph{Mechanized Symbolic Protocol Analysis} %
Over the years, several works have focused on the mechanized security of cryptographic protocols in the symbolic security model defined by Dolev and Yao~\cite{dolev-yao}.
In his seminal work, Lawrence C. Paulson~\cite{Paulson98} developed a rigorous technique for verifying cryptographic protocols within the Isabelle/HOL proof assistant~\cite{isabelle}.
The approach is based on representing protocols as (potentially unbounded) sets of execution traces, capturing the messages exchanged among protocol participants and with the adversary.
Leveraging the expressiveness of Isabelle, particularly for inductive reasoning, the approach enables proving, for instance, that a certain communication event always follows another event, under the assumption that data remains secret.
The authors demonstrate the effectiveness of the approach by verifying the NSL protocol and, in a later work, the Kerberos authentication protocol~\cite{BellaPaulson98}.
Despite its expressiveness, the approach offers limited automation: while Isabelle provides general-purpose tools (e.g., sledgehammer~\cite{blanchetteExtendingSledgehammerSMT2013}), proofs remain largely manual due to the need for protocol-specific reasoning and the challenges of automating inductive arguments in general-purpose proof search.

Busi et al.~\cite{Busi25} follow a similar approach in their formalization of strand spaces.
Their \mStrandsRocq library encodes the inductive proof technique developed by Fábrega et al.~\cite{FHG99}, enabling symbolic proofs of secrecy and agreement within the Rocq proof assistant.
\mStrandsRocq provides a set of strand-specific automated tactics for implementing case analysis over strands, automatically solving trivial goals.
While this automation is specialized and supports proof reuse, it is constrained by a protocol representation that remains close to the original strand space model.
This design choice limits the degree of achievable automation compared to approaches following the LAC principle~\cite{Ricketts14}, where the language can be deliberately restricted to enable more effective automated reasoning.

The \mDYstar library~\cite{dy-star} aims to provide a different workflow, where protocol implementations can be automatically verified by leveraging the SMT-based automation of the \mfstar language.
Protocols are modeled as imperative procedures using effectful primitives to send and receive messages from the network, store state, and perform cryptographic operations.
These primitives manipulate a global trace of events for verification and can be swapped for concrete implementations for execution.
This fully automated approach, however, suffers from several shortcomings.
First, the code needs to be annotated with \code{assert} commands to guide the SMT solver toward a solution.
These proof annotations obscure the flow of the protocol, mixing specification and technical proof-related details.
Second, \mfstar offers limited tool support for writing and refining these assertions, making it more difficult to guide the verification process interactively compared to proof assistants that provide more direct feedback.
To address the limitations of this automated (intrinsic) proof style, Théophile Wallez developed an alternative version of \mDYstar focusing on explicit extrinsic proofs~\cite{wallezDYUnchainedNow2026,wallez25}.
Compared to automated SMT-based reasoning, extrinsic proofs require a less automated workflow, similar to more traditional interactive theorem proving.
This shift, however, further highlights the limitations of \mfstar with respect to interactive, tactic-based proof development, and motivates our choice for the \mlean proof assistant.
This work presents a further extension of \mDYstar to support more general labels, capable of modeling arbitrary trace properties.
Compared to our implementation, however, their labeling system does not support the XOR operator nor the automatic labeling of non-atomic values described in \autoref{sec:releasable}.

\bfparagraph{Automated Symbolic Protocol Analysis} %
In parallel with the above efforts on the mechanized analysis of protocols, fully automated solvers have been developed and refined over the years.
Notably, Bruno Blanchet introduced \mproverif~\cite{proverif}, one of the earliest and most widely used tools supporting so-called push-button verification of protocols expressed in the applied $\pi$-calculus~\cite{AF01}.

\mproverif translates $\pi$-calculus processes into Horn clauses and employs a specialized resolution-based algorithm to check secrecy and correspondence properties.
A comparable approach is taken by Simon Meier in the \mtamarin prover~\cite{tamarin}.
In \mtamarin, protocols are specified as rewriting rules, and the solver explores the resulting transition system over multisets of terms, enabling the verification of secrecy and correspondence properties.
Both tools support the analysis of protocols with an unbounded number of sessions and messages.
However, they require reasoning about the full protocol execution, lacking mechanisms to analyze sub-protocols in isolation and to compose verification results.
In subsequent work, \mproverif has been extended to support observational equivalence~\cite{chevalProvingMoreObservational2013}, a limited form of induction~\cite{blanchetProVerifLemmasInduction2022} over the length of execution traces, and even support for XOR~\cite{chevalAutomaticVerificationFinite2025}.
Additionally, various techniques have been proposed to integrate support for XOR operations into these automated verifiers, such as reducing Horn theories with XOR to the XOR-free case to enable analysis in \mproverif~\cite{Kuesters08}, or extending the \mtamarin prover to support stateful cryptographic protocols with XOR~\cite{DHRS2018}.
Attempts were also made to provide a hammer-like approach combining all these tools in the form of \msapic~\cite{sapic+}.
While all these approaches move automated verifiers toward greater flexibility,
they offer only limited support for interactive proof development when automation alone is insufficient.

\bfparagraph{Blockchain Protocol Verification}
The emergence of blockchain technologies introduced new classes of protocols and security properties to verify, prompting a rapid response from the verification community.
Significant effort has been devoted to formalizing the foundations of blockchain systems.
Rupi\'c \etal~\cite{rupicMechanizedFormalModel2020} provide a mechanized formalization of Bitcoin's consensus in \mrocq,
while Annenkov \etal~\cite{annenkovConCertSmartContract2020} formalize the smart contract execution layer.
At the contract level,
Andrychowicz \etal~\cite{andrychowiczModelingBitcoinContracts2014} model contracts using \textsc{UPPAAL}~\cite{uppaal}, and Jacquot and Donnet develop \textsc{CHAUSSETTE}~\cite{jacquotCHAUSSETTESymbolicVerification2023}, a tool that uses symbolic execution to uncover vulnerabilities in Bitcoin scripts.
Our work, instead, focuses on modeling layer-2 (off-chain) protocols~\cite{ln-main}, specifically payment channels and the Lightning Network.

The Lightning Network has attracted substantial research interest across several aspects.
This includes game-theoretic analyses such as \mcheckmate~\cite{bruggerCheckMateAutomatedGameTheoretic2023} and the work of Rain \etal~\cite{rainGameTheoreticSecurityAnalysis2023}.
Meanwhile, in the symbolic model, Boyd \etal~\cite{boydBlockchainModelTamarin2020a} leverage \mtamarin to formally
 analyze Hash Time-Lock Contracts (HTLCs), and Barthe \etal~\cite{BDLMR22} develop \mtidy to
 extend \mtamarin with support for timed cryptographic protocols.
Additionally, H\"uttel and Starove\v{s}ki~\cite{huttelSecrecyAuthenticityProperties2020,huttelKeyAgreementLightning2022} verified secrecy and authenticity properties of specific Lightning subprotocols using \mproverif.
The works closest to ours in scope are the recent verifications of the full Lightning Network protocol.
These include the TLA+ model by Grundmann and Hartenstein~\cite{Grundmann26} and the \mwhy~\cite{why3} formalization by Fabia\'{n}ski \etal~\cite{fabianskiFormallyVerifiedLightning2026}.
While these are impressive verification achievements, they focus on the monolithic verification of a single protocol, rather than providing a reusable framework.
In contrast, we focus on the usability and generality of \leandy.
Our formalization of payment channels in \leandy demonstrates that our library is expressive enough for this class of protocols, providing a general framework for the verification of safety and liveness properties of complex protocols.

\section{Conclusion}
In this paper, we presented \leandy, a library for symbolic protocol verification in the Lean proof assistant that combines type-based reasoning with inductive trace-based proofs.
The framework builds on the design of \mDYstar and supports mutable state, dynamic compromise, and conditional secrecy properties for protocols using XOR, including chained release patterns. Following the LAC design principle, \leandy provides protocol-specific automation while still allowing users to complement it with interactive proofs when more expressivity is required.

We developed a mechanized model of SegWit-style blockchain primitives in \leandy, capturing Bitcoin-like transaction structures, witness-based scripts, and timelocks. We showed the expressiveness of the framework by developing an in-depth formalization of payment channels, verifying punishment mechanisms and properties that depend on chain liveness.
As future work, we plan to build on this payment-channel development toward a formalization of the full Lightning Network protocol.

\bibliographystyle{ACM-Reference-Format}
\bibliography{bibliography.bib}

\appendix

\end{document}